\documentclass[aps,12pt,prd,superscriptaddress,preprintnumbers, 
	amssymb,
	amsmath,
	notitlepage,
	longbibliography,
	nofootinbib]{revtex4-1}
\usepackage[colorlinks=true, 
	linkcolor=blue, 
	citecolor=blue, 
	urlcolor=blue, 
	linktocpage=true]{hyperref}
\usepackage[utf8]{inputenc}
\usepackage[english]{babel}
\usepackage{lmodern}
\usepackage[protrusion=true, expansion=true]{microtype}
\usepackage[dvipsnames]{xcolor}
\usepackage{graphicx}
\usepackage{mathtools}

\usepackage[
]{todonotes}

\usepackage[framemethod=TikZ]{mdframed}
\usepackage{lipsum}

\mdfdefinestyle{MyFrame}{
linecolor=blue,
outerlinewidth=2pt,
roundcorner=20pt,
innertopmargin=\baselineskip,
innerbottommargin=\baselineskip,
innerrightmargin=20pt,
innerleftmargin=20pt,
backgroundcolor=gray!20!white}

\newcommand{\lp}{\ensuremath{\left(}}
\newcommand{\rp}{\ensuremath{\right)}}
\newcommand{\s}{\nobreak\hspace{.08em plus .04em}}

\newcommand{\e}{\ensuremath{\mathrm{e}}}

\newcommand{\real}{\mathbb{R}}

\newcommand{\be}{\begin{equation}}
\newcommand{\ee}{\end{equation}}
\newcommand{\beal}{\begin{aligned}}
\newcommand{\eeal}{\end{aligned}}
\newcommand\bea {\begin{eqnarray}}
\newcommand\eea {\end{eqnarray}}
\newcommand{\bec}{\begin{cases}}
\newcommand{\eec}{\end{cases}}
\newcommand{\bei}{\begin{itemize}}
\newcommand{\eei}{\end{itemize}}
\newcommand{\bee}{\begin{enumerate}}
\newcommand{\eee}{\end{enumerate}}

\definecolor{remFcol}{RGB}{0,140,220}

\begin{document}

\preprint{DCPT-18/27}

\title{Negative modes of Coleman-de Luccia and black hole bubbles}
\date{\today} 

\author{Ruth Gregory}
\email{r.a.w.gregory@durham.ac.uk}
\affiliation{Centre for Particle Theory, Durham University,
South Road, Durham, DH1 3LE, UK}
\affiliation{Perimeter Institute, 31 Caroline Street North, Waterloo, 
ON, N2L 2Y5, Canada}
\author{Katie M. Marshall}
\email{k.marshall6@newcastle.ac.uk}
\affiliation{School of Mathematics, Statistics and Physics, Newcastle University, 
Newcastle Upon Tyne, NE1 7RU, UK}
\author{Florent Michel}
\email{florent.c.michel@durham.ac.uk}
\affiliation{Centre for Particle Theory, Durham University,
South Road, Durham, DH1 3LE, UK}
\author{Ian G. Moss}
\email{ian.moss@newcastle.ac.uk}
\affiliation{School of Mathematics, Statistics and Physics, Newcastle University, 
Newcastle Upon Tyne, NE1 7RU, UK}

\begin{abstract} 
We study the negative modes of gravitational instantons representing 
vacuum decay in asymptotically flat space-time.
We consider two different vacuum decay scenarios: the 
Coleman-de Luccia $\mathrm{O}(4)$-symmetric bubble, and 
$\mathrm{O}(3) \times \mathbb{R}$ instantons with a
static black hole. In spite of the similarities between the models, we 
find qualitatively different behaviours.
In the $\mathrm{O}(4)$-symmetric case, the number of negative modes 
is known to be either one or infinite, depending on the sign of the kinetic term in 
the quadratic action. In contrast, solving the mode equation numerically
for the static black hole instanton, we find only one negative mode with the
kinetic term always positive outside the event horizon. The absence of
additional negative modes supports the interpretation of these solutions 
as giving the tunnelling rate 
for false vacuum decay seeded by microscopic black holes. 
\end{abstract}

\maketitle

\section{Introduction}

False vacuum decay through the nucleation of true vacuum bubbles
has many important applications ranging from early universe 
phase transitions to stability of the Higgs vacuum. 
The process has an elegant description in terms of Euclidean solutions to the 
underlying field equations that extend the bubbles into imaginary time  
\cite{Coleman:1977,Callan:1977}. Path integral methods give vacuum 
decay rates which depend on the Euclidean action of the bubble solutions 
and the eigenvalues of perturbative modes on the bubble background.

A crucial feature of the bubble nucleation picture is the existence of a single
negative mode of field perturbations, which corresponds physically
to scaling the size of the bubble up or down. In the analysis of the 
vacuum decay process, the square root of this negative mode provides
an imaginary part to the energy of the false vacuum state, which in turn corresponds to
a decay rate. The analysis would fail if, for example, we have two negative
modes in which case the net contribution to the energy becomes real.
Fortunately, it is known that the basic picture with a single bubble
has just the single negative mode \cite{Coleman:1988}.

Coleman and de Luccia \cite{Coleman:1980aw} were the first people to 
extend the basic formalism of vacuum decay to include the effects of 
gravitational back-reaction in the bubble solutions,
producing a type of gravitational instanton.
The negative modes of the Coleman-de Luccia instanton that represents
vacuum decay in de Sitter space have been studied by several authors
\cite{Lavrelashvili:1985vn,Tanaka:1992zw,Garriga:1993fh,Gratton:1999ya,
Lavrelashvili:1999sr,Tanaka:1999pj,Gratton:2000fj,Khvedelidze:2000cp,
Dunne:2006bt,Lee:2014uza,Koehn:2015hga},
and there is now compelling evidence that the bounce solutions can have 
either one or infinitely many negative modes, depending on the potential. 
The existence of many negative modes seems to be associated
with situations where the bounce solution is comparable in size to the
cosmological horizon \cite{Lee:2014uza}.

The past few years have seen a resurgence of interest in applications
of vacuum decay to the standard model Higgs field \cite{Krive:1977,
Cabibbo:1979ay,Polizer:1979,Isidori:2001bm,Rajantie:2016hkj,
Andreassen:2017rzq,Branchina:2018xdh}. 
Depending on the values of the Higgs and top quark masses, the quantum corrected
Higgs potential can decrease at large field values and destabilise
the present day 246 GeV minimum. The scale at which an instability sets in, $\Lambda$,
is very sensitive to particle physics parameters and possible new physics,
with typical ranges $10^{10}-10^{18}\,{\rm GeV}$ leading to a metastable
false vacuum \cite{Degrassi:2012ry,Buttazzo:2013uya,Blum:2015rpa}.
Vacuum decay rates are strongly exponentially suppressed, but
recently the possibility of black holes seeding vacuum decay has
been considered \cite{Gregory:2013hja,Burda:2015isa,Burda:2015yfa,
Burda:2016mou,Gregory:2016xix,
Tetradis:2016vqb,Chen:2017suz,Mukaida:2017bgd} 
and the decay in this case is very rapid. Its implications for 
early cosmology have been investigated in~\cite{Gorbunov:2017fhq}. 
In parallel, it was shown in~\cite{Chen:2018aij} that Hawking radiation 
can be described by a family of instantons. 

Similar ideas have been discussed in the context of eternal inflation, 
see for instance~\cite{Aguirre:2005nt}, as well as~%
\cite{Aguirre:2006ap,Bousso:2006am} in which the limit of vanishing 
cosmological constant in the false vacuum phase is studied in details. 
It is found, in particular, that this limit is continuous, contrary to 
what was previously conjectured.

The negative mode problem has so far only been investigated numerically
for vacuum decay in asymptotically de Sitter spacetimes. In this paper 
we give the first analysis of negative modes for the asymptotically flat 
bounces that are relevant for decay of the Higgs vacuum. We look at two 
different Higgs vacuum decay scenarios, vacuum decay in empty space and 
vacuum decay seeded by black holes. Vacuum decay rates with gravitational 
back-reaction in empty space have been examined by 
\cite{Isidori:2007vm,Rajantie:2016hkj,Salvio:2016mvj,Rajantie:2017ajw}.
The gravitational back reaction is significant when $\Lambda$ approaches the
Planck scale, as might be expected. Non-minimal coupling of the Higgs field to gravity
can have a significant effect on the decay process, and so we include this
possibility on our negative mode analysis.

For decay in empty space, we find numerically that there is 
either a single negative mode, or infinitely many as in the de Sitter case. 
The emergence of the infinite tower of negative modes is related to a change 
in sign for the kinetic terms in the action of the perturbations. This is also seen 
in the asymptotically de Sitter case. We have used an approach where the 
gravitational constraints are used to eliminate extraneous gauge degrees of 
freedom. Our approach is therefore free of gauge artefacts, and gives similar 
equations to those in Ref \cite{Lee:2014uza},
where a gauge invariant parameterisation was used.

The second scenario we have investigated is the case where vacuum decay
is enhanced by the presence of a microscopic black hole left over from
the early universe. The black hole
acts as a nucleation seed and greatly enhances the vacuum decay rate.
This effect was investigated initially for vacuum decay in de Sitter space 
\cite {Gregory:2013hja},
and later for more general scenarios including asymptotically flat space 
\cite{Burda:2015isa,Burda:2015yfa,Burda:2016mou}. 
In all cases, the dominant decay process is one with static O(3) symmetric 
bubbles. We shall give the results of a numerical analysis of the negative 
modes for vacuum decay with an asymptotically flat black
hole nucleation seed. In this case we find only one
negative mode, and the kinetic term in the action of the perturbations
is always positive. We conclude from this that vacuum decay seeded by
black holes most likely always has a consistent formulation in terms of bounce
solutions.

\section{Tunnelling and negative modes}

We consider decay of the false vacuum state of a scalar field $\phi$ 
with potential $V(\phi)$. Tunnelling from the false vacuum is represented 
in the path integral formalism by bounce solutions 
$\phi_b$ to the scalar field equations, with imaginary time coordinate 
$\tau$~\cite{Coleman:1977}. Boundary conditions  
are $\phi_b\to \phi_{\rm fv}$ when $\tau\to\pm\infty$
and at spatial infinity $|{\bf x}|\to\infty$, where $\phi_{\rm fv}$ is the 
value of the field at the false vacuum. 
The tunnelling exponent for a bounce solution is related to the change in 
Euclidean action by $B=S_E[\phi_b]-S_E[\phi_{\rm fv}]$, where
\begin{equation}
B=\int_{-\infty}^\infty d\tau\int d^3x \left(\frac12(\partial_\tau\phi_b)^2+\frac12(\boldsymbol{\nabla}\phi_b)^2
+V(\phi_b) 
\right).
\end{equation}
Given reasonable conditions on the potential, it has been shown
\cite{Coleman:1977th} that there is a bubble solution
with $O(4)$ symmetry that has the smallest action, and hence the 
largest tunnelling rate, compared to other bounce solutions. Furthermore, 
this solution has exactly one negative mode~\cite{Callan:1977}, and is 
therefore a saddle point of the Euclidean action.

Evaluating the path integral for a single bubble solution gives a contribution 
to the vacuum decay amplitude of the form
\begin{equation}
I_{\rm bubble}\approx \frac12i\Omega T\left|
{{\rm det}'\,S_E''[\phi_b]\over {\rm det}\,S_E''[\phi_{\rm fv}]}
\right|^{-1/2} {B^2\over 4\pi^2}\,\e^{-B}\,
I_{\rm fv},
\end{equation}
where $S_E''$ denotes the second functional derivative of the Euclidean action, 
and det$'$ denotes omission of zero modes from the determinant. The zero modes
give factors $\Omega$ and $T$ for the total volume and time period, along with 
a Jacobian factor $B^2/4\pi^2$. The factor $i$ arises from the negative mode. 
This would become $i^n$ if there were $n$ negative modes. The vacuum decay 
rate $\Gamma$ can be calculated by summing multiple bubble amplitudes, 
and the result is \cite{Coleman:1977th,Callan:1977}
\begin{equation}
\Gamma \approx \left|
{{\rm det}'\,S_E''[\phi_b]\over {\rm det}\,S_E''[\phi_{\rm fv}]}
\right|^{-1/2}\,{B^2\over 4\pi^2}\,e^{-B}.
\end{equation}

The negative mode can be explained easily in the thin-wall limit, when the 
bubble solution consists of a true vacuum region $\phi_{\rm tv}$ surrounded 
by a relatively narrow wall where the field transitions to the false vacuum. 
This approximation is valid when the difference in energy $\varepsilon$ of the
true and false vacua is small compared to a combination of barrier height and width.
The field is represented by a bubble Ansatz of the form 
$\phi=\phi(r;R)\simeq \phi_0(r-R)$, where $\phi_0(x)$ solves the
`planar' domain wall equation 
\be
\phi'' \approx \frac{\partial V}{\partial \phi}\,.
\label{kinkeom}
\ee
Provided the bubble is large compared to the wall thickness this is an excellent
approximation, and allows us to integrate the tunnelling exponent in terms
of the bubble radius $R$,
\begin{equation}
B(R)=2\pi^2\sigma R^3-\frac12\pi^2\varepsilon R^4.
\end{equation}
Here, $\sigma$ is the action per unit area of the bubble wall, which can be
found in terms of an integral of the potential
$V(\phi)$ by
\begin{equation}
\sigma=\int_{\phi_{\rm fv}}^{\phi_{\rm tv}}\,|2\Delta V(\phi)|^{1/2}d\phi
\end{equation}
using $\frac12\phi_0^{\prime2} = \Delta V$ from \eqref{kinkeom}.
The bubble solution is given by the extremum at the radius $R_b
=R_0\equiv3\sigma/\varepsilon$, where $B$ has a maximum. 

The negative mode corresponds to changes in $\phi$ that increase 
or decrease the radius of the bubble solution,
\begin{equation}
\delta\phi={d\phi\over dR}\delta R.
\end{equation}
The overall change in $B$ is related to the negative eigenvalue $\lambda_0$ by,
\begin{equation}
\delta B\approx \frac12 B''(R)
\delta R^2\approx\frac12 \left\lVert \delta\phi\right\rVert^2\lambda_0,
\end{equation}
where the norm of a function $f(x)$ is defined by
\begin{equation}
\left\lVert f\right\rVert^2=\int f(x)^2\,d^4x.
\end{equation}
We therefore have a simple formula for the negative mode in the thin-wall approximation,
\begin{equation}
\lambda_0\approx\left.{B(R)''\over ||d\phi/dR||^2}\right|_{R=R_b},
\label{thinwallev}
\end{equation}
This can be taken further using our approximation for the bubble wall profile, 
since $d\phi_0/dR=-\phi_0'$, hence $||d\phi/dR||^2=||\phi'||^2
\approx2\pi^2\sigma R^3$, and we have
\begin{equation}
\lambda_0\approx -{3\over R_b^2}.
\end{equation}
The approximation is valid when the thickness of the wall is small 
compared to the bubble radius,
which translates to $\varepsilon\ll9\sigma^2/\phi_{\rm tv}^2$. 

Now we turn to bubble solutions with gravitational back-reaction. 
These can be found by extremising the Einstein-scalar action,
\begin{equation} \label{eq:gaction}
S_E = \int \lp - \frac{{\cal R}}{16 \s \pi \s G} + \frac12 \lp \partial \phi \rp^2
+ V(\phi) \rp \sqrt{g} \s d^4 x,
\end{equation}
where ${\cal R}$ is the Ricci scalar. Bubble solutions with $O(4)$ symmetry 
can be described by a `radial' solution of scalar field, $\phi(r)$, and geometry:
\be
ds^2 = dr^2 + a^2(r) d\Omega_{I\!I\!I}^2
\ee
where $\phi$ and $a$ tend towards the true vacuum form as $r,a(r)\to0$, 
and the false vacuum form for large $r$.
We take a leap of faith in assuming that the vacuum decay exponent for 
a single bubble is still given by the difference in Euclidean 
action between the bubble solution and the false vacuum. 
There are two distinct scenarios, depending on whether the Euclidean
metric is compact or infinite. In the compact case, the scalar field never 
quite reaches the false vacuum value outside the bubble, but regularity 
conditions on the metric at the two points where $a=0$ restrict the possible 
bubble solutions. In the infinite case, the scalar field asymptotically approaches 
the false vacuum value as $a\to\infty$.
In this case we require that the Euclidean metric approaches the same form 
for the bounce and for false vacuum to ensure that the tunnelling exponent 
$B=S_E[\phi_b]-S_E[\phi_{\rm fv}]$ is finite. (Note that adding boundary terms 
to the Einstein-scalar action is unnecessary as
these cancel out when evaluating $B$.)
\begin{figure}
\includegraphics[width=0.49\linewidth]{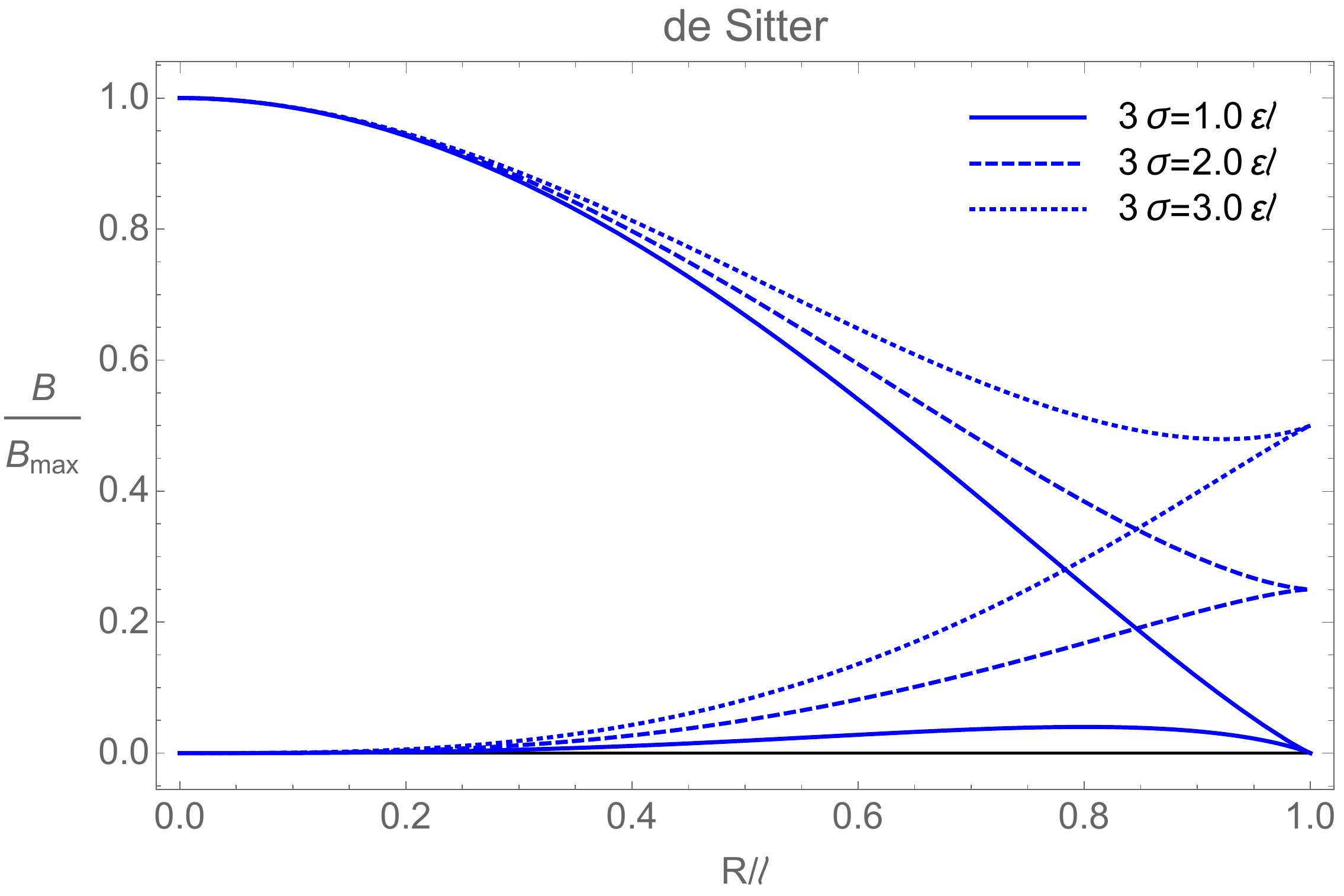}
\includegraphics[width=0.49\linewidth]{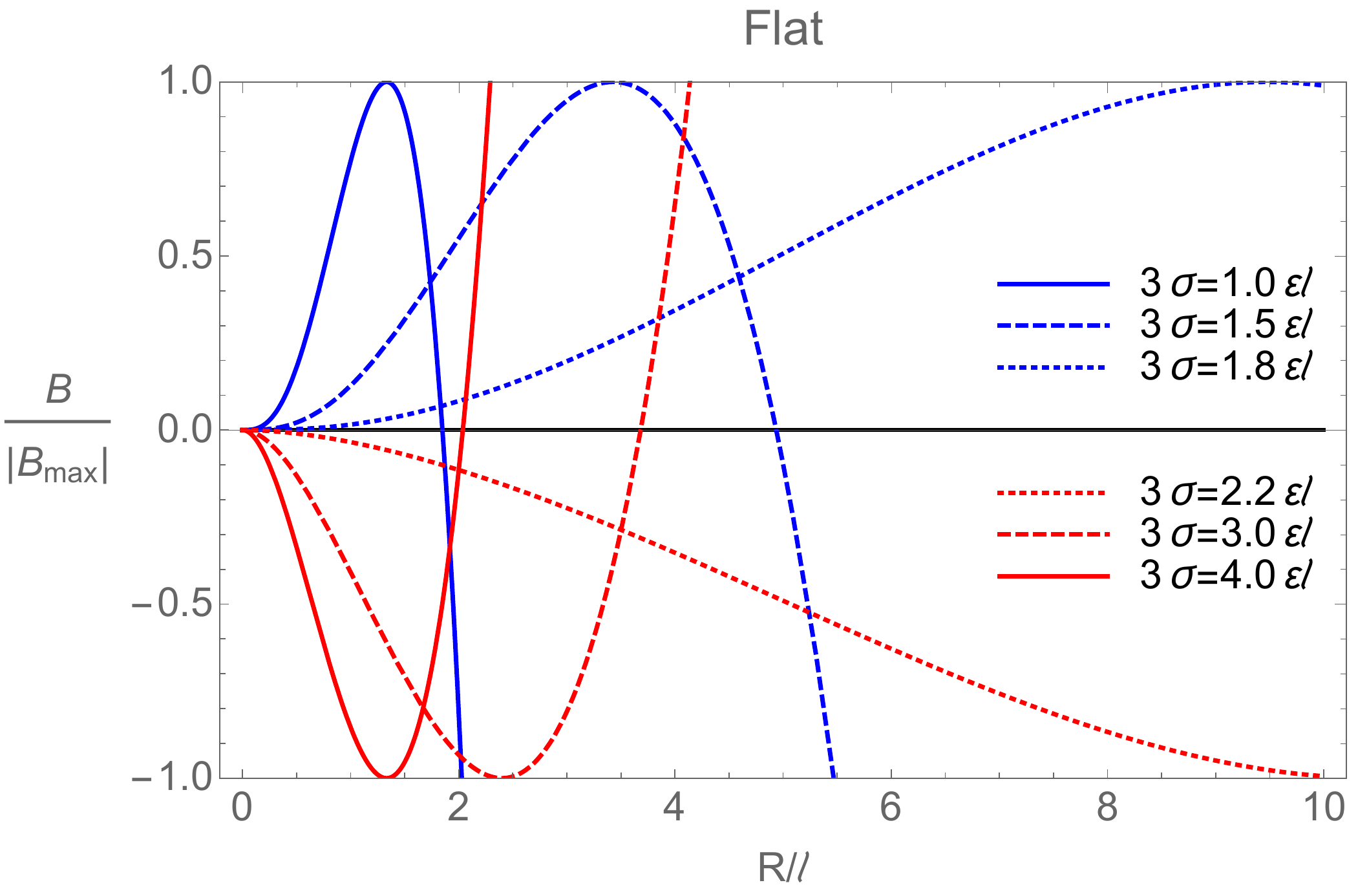}
\caption{
Left panel: The tunnelling exponent $B(R)$ for a thin-wall bubble of flat 
vacuum in de Sitter space. The large and small bubble exponents are superposed. 
Right panel: The tunnelling exponent $B(R)$ for a 
thin-wall bubble of anti-de Sitter vacuum in flat space.
} 
\label{fig:thinwall}
\end{figure}

Following Coleman and de Luccia, we gain some insight into the $O(4)$ 
bubble solutions by taking a thin-wall limit. The thin wall approximation
assumes that the scalar varies rapidly between its false and true vacuum
over a region $w \ll R$, where $R$ as before represents the 
size of the bubble radius that we take as the areal radius: the value of the scale 
factor at the centre of the bubble wall. The thin wall approximation is
valid provided that the local spacetime curvature induced within 
the wall remains below the Planck scale $w\sigma\ll M_p^2$ 
\cite{Garfinkle:1989mv}. 
(In this work we use the reduced Planck mass, defined by $M_p^2 = 1 / (8 \pi G)$.)

We shall see in the following section that the curved-space 
bubble solutions can be represented by the form $\phi=\phi(a;R)$ where 
$\phi\approx\phi_0(r-r_b)$ for the thin wall, with $r_b$ the coordinate 
location of the bubble centre: $a(r_b)=R$. We then approximate the scale
factor by a piecewise differentiable function 
\be
a(r) = a_{{\rm tv}}(r) \,\Theta[r_b-r] + a_{\rm fv}(r) \, \Theta[r-r_b]\,,
\label{thinwalla}
\ee
where $a_{\rm tv}(r_b) = a_{\rm fv}(r_b) = R$, and compute the difference in 
action between the bubble and false vacuum configurations for this Ansatz.
The Ansatz also allows us to estimate the negative eigenvalue as before, but
with the norm calculated using an appropriate curved space measure.

Considering first the compact case,
we take the false vacuum to have positive energy $\varepsilon$,
and the true vacuum to have zero energy. This represents the decay of the false 
vacuum from a de Sitter universe into flat space, thus $a_{\rm tv}=r$
and $a_{\rm fv}=\ell \sin((r-r_0)/\ell)$ in \eqref{thinwalla},
where $\ell=\sqrt{3/(8\pi G\varepsilon)}$ is the de Sitter radius
and $r_0$ is a introduced to satisfy $a_{\rm tv}(r_b) = a_{\rm fv}(r_b) = R$.
The tunnelling exponent can
be directly calculated as (see also \cite{Coleman:1980aw})
\begin{equation}
B(R)=\frac43\pi^2\varepsilon\ell^4\left\{1\mp(1- R^2/\ell^2)^{3/2}\right\} 
- 2\pi^2\varepsilon\ell^2 R^2 +2\pi^2\sigma R^3,
\label{actionds}
\end{equation}
and is plotted in figure \ref{fig:thinwall}.
The upper sign applies when the false vacuum region is larger than a hemisphere,
and the lower sign applies when the false vacuum covers less than a hemisphere.
In the first case, the true vacuum bubble encloses a smaller volume than the false 
vacuum region, and in the second case the true and false vacuum regions have 
a similar volume. Following Ref. \cite{Lee:2014uza}, we refer to these as the small 
bubble and large bubble situations.

The exponent \eqref{actionds} has one extremum $R_b$ away from the origin,
\begin{equation}
R_b={R_0\over 1+(R_0/2\ell)^2},
\end{equation}
where $R_0=3\sigma/\varepsilon$ is the bubble radius without the gravitational 
back reaction. Bubble solutions always exist, but the extremum becomes a
minimum when $3\sigma/\varepsilon>2\ell$. The thin-wall approximation therefore 
predicts the disappearance of the negative mode, and we can estimate the
value of the mode in a similar way to the probe case. Since the bubble wall
is determined by $r=R$, and the geometry inside the bubble is flat, we
find that the eigenvalue is well approximated here by the flat space value
\eqref{thinwallev}. Numerical investigations have shown that new sets of 
spherically symmetric negative modes start to appear \cite{Lavrelashvili:1985vn,
Tanaka:1992zw,Lavrelashvili:1999sr,Lee:2014uza,Koehn:2015hga}.
The first set are fluctuations localised near the bubble wall,
called `wall modes' in Ref \cite{Lee:2014uza}. The second set are localised 
near the maximum radius of the instanton in the `large bubble' case.

In the non-compact case, the true vacuum has negative energy $-\varepsilon$ 
and the false vacuum has zero energy. This represents vacuum decay from 
flat space to anti de Sitter space, and we have $a_{\rm tv}=\ell \sinh r/\ell$,
$a_{\rm fv}=r + (R-r_b)$ in \eqref{thinwalla}. This time one finds
\begin{equation}
B(R)=\frac43\pi^2\varepsilon\ell^4\left\{1-(1+ R^2/\ell^2)^{3/2}\right\} 
\pm 2\pi^2\varepsilon\ell^2 R^2+2\pi^2\sigma R^3,
\label{bads}
\end{equation}
plotted in the right panel of figure \ref{fig:thinwall}.
The upper sign corresponds to $3\sigma/\varepsilon<2\ell$,
and the bubble matches an interior of AdS to an \emph{exterior} of 
an $S^3$ in $\real^4$, i.e.\ a conventional bubble. 
For $3\sigma/\varepsilon>2\ell$ we can still find a solution, provided we 
match the interior of an AdS sphere to an \emph{interior} of a three-sphere
in $\real^4$: clearly this does not have an intuitive interpretation as a
vacuum decay bubble, and is similar to the situation of dS tunneling
above, where the false vacuum covers less than a hemisphere of dS. We note
simply that these solutions do not have a negative mode, hence are not
tunneling instantons, and do not consider them further. For 
$3\sigma/\varepsilon<2\ell$ the bubble has radius
\cite{Coleman:1980aw}
\begin{equation}
R_b={R_0\over 1-(R_0/2\ell)^2}.
\label{rads}
\end{equation}
Whenever a bubble solution exists the extremum is always a maximum 
and the negative mode we had originally should remain.
This time, in our estimate of the negative eigenvalue,
we note $R=\ell \sinh r_b/\ell$, hence
\begin{equation}
\left\lVert {d\phi\over dR}\right\rVert^2=
\left\lVert {d\phi\over dr_b}\right\rVert^2\left({dr_b\over dR}\right)^2
=2\pi^2\sigma R^3(1+ R^2/\ell^2)^{-1}
\end{equation}
We can substitute this into the general formula (\ref{thinwallev}), 
with the exponent $B(R)$ from (\ref{bads}), and evaluate the result at the 
bubble radius $R_b$ from (\ref{rads}), to get
\begin{equation}
\lambda_0\approx -{3\over R_0^2}\left\{1-\left({R_0\over 2\ell}\right)^4\right\},
\label{evapprox}
\end{equation}
where $R_0=3\sigma/\varepsilon$ as before. This formula will be used to check 
the results of the numerical study of the negative modes presented 
in the following section.

\section{Vacuum decay through {O(4)}-symmetric bubbles}
\label{sec:O4}

\subsection{Model and field equations}
\label{sub:O4:model}

In order to consider a wide variety of models of interest to Higgs cosmology, 
we generalise the gravitational action (\ref{eq:gaction}) to include non-minimal
coupling between the scalar field and gravity, 
\begin{equation}
S = \int \lp - \frac{\hat{R}}{16 \s \pi \s G} + \frac{\xi}{2} \s \hat{R} \s \phi^2 
+ \frac{\hat{g}^{\mu \nu}}{2} \s (\partial_\mu \phi) \s (\partial_\nu \phi) 
+ V(\phi) \rp \sqrt{\hat{g}} \s d^4 x,
\end{equation}
where $\xi$ is a non-minimal coupling coefficient and hats denote the 
choice of metric commonly referred to as the Jordan frame. 
We consider potentials such that $V(0) = V'(0) = 0$, $V''(0) > 0$, and 
assume $V$ takes negative values 
in some interval of $\phi$ so that the bubble solutions will be non-compact.
To find numerical solutions and study their perturbations, it is convenient 
to go to the Einstein frame by rescaling the metric:
\begin{equation}
g_{\mu \nu} = \lp 1 - 8 \s \pi \s G \s \xi \s \phi^2 \rp \hat{g}_{\mu \nu},
\end{equation}
(for an analysis of solutions in the Jordan frame see 
\cite{Rajantie:2016hkj,Rajantie:2017ajw}).
The action becomes
\begin{equation} \label{eq:O4:offsaction}
S = \int \lp - \frac{R}{16 \s \pi \s G} 
+ \frac{f(\phi)^2}{2} \s \lp \partial_\mu \phi \rp \lp \partial^\mu \phi \rp + 
W(\phi) \rp \sqrt{g} \s d^4 x,
\end{equation}
where 
\begin{equation}
f(\phi) = \frac{\sqrt{1 - 8 \s \pi \s G \s \xi \lp 1 
- 6 \s \xi \rp \phi^2}}{1 - 8 \s \pi \s G \s \xi \s \phi^2}
\end{equation}
and the modified potential is
\begin{equation} \label{eq:O4:W}
W(\phi) = \frac{V(\phi)}{\lp 1 - 8 \s \pi \s G \s \xi \s \phi^2 \rp^2}.
\end{equation}
In all the cases we will consider, $f(\phi)$ remains strictly positive. 
We look for $\mathrm{O}(4)$-symmetric solutions, and change slightly
the form of our metric to add a lapse function:
\begin{equation} \label{eq:O4:met}
ds^2 = N(\rho)^2 \s d \rho^2 + a(\rho)^2 \s d \Omega_{I\!I\!I}^2,
\end{equation}
The lapse function $N$ allows us to recover the full 
set of Einstein equations from extremization of the action, 
which will be convenient when deriving the eigenvalue equation. 
Substituting in the form of the metric \eqref{eq:O4:met}, and integrating
out over the angular variables, we obtain
\begin{equation} \label{eq:O4:ac1}
S = 2 \s \pi^2 \int \left [ \frac{f(\phi)^2}{2 \s N^2} \s \phi^{\prime 2} 
+ W(\phi) - \frac{3}{8 \s \pi \s G} \s \lp \frac{1}{a^2} + 
\lp \frac{a'}{a \s N} \rp^2 \rp \right ] a^3 \s N \s d \rho,
\end{equation}
Variation with respect to $\phi$ and $N$ give the system of equations:
\begin{align}
& f(\phi) \s \lp f(\phi) \s \frac{a^3}{N} \s \phi' \rp' = N \s a^3 \s W', \label{eq:O4:phi} \\
& \frac{a^{\prime2}}{N^2} = 1 + \frac{8 \s \pi \s G}{3} \s a^2 \s \lp 
\frac{f(\phi)^2}{2 \s N^2} \s \phi^{\prime 2} - W(\phi) \rp. 
\label{eq:O4:a}
\end{align}
Variation with respect to $a$ gives a Bianchi Identity\footnote{Using 
Eq.~\eqref{eq:O4:phi}, 
it is equivalent to the derivative of Eq.~\eqref{eq:O4:a}.}.
The system~(\ref{eq:O4:phi},\ref{eq:O4:a}) can also be obtained from the 
full set of Einstein equations after eliminating redundancies, showing that 
there is no independent constraint.
For boundary conditions, we look for asymptotically flat instantons, 
with $\phi(\infty) = \phi_{FV}$ and $a(\rho) \sim \rho$ as $\rho \to \infty$. 
We choose to place the centre of the instanton at $\rho=0$, where $a(0)=0$
and for regularity at the origin we must have $\phi'(0) = 0$.  
Equation~\eqref{eq:O4:a} can be rewritten as: 
\begin{equation}
\frac{1 - 8 \s \pi \s G \s a^2 \s W(\phi) / 3}
{1 - 4 \s \pi \s G \s a^2 \s (\partial_a \phi)^2 / 3} = \frac{a^{\prime2}}{N^2}.
\end{equation}
This shows that the left-hand side, which will play an important role in the 
following, is always non-negative, and cannot vanish if $a$ is 
strictly 
monotonic. 

The lapse function $N(\rho)$ represents some of the freedom we have to 
choose the coordinate gauge. We will  focus on instantons where $a$ is a 
strictly increasing 
function of the distance to the center of the bubble,
which allows us to choose $a$ as radial coordinate. 
Setting $\rho = a$, the action~\eqref{eq:O4:ac1} becomes
\begin{equation}\label{eq:O4:ac3}
S = 2 \s \pi^2 \s \int_0^\infty N \s a^3 \s \lp \frac{ f(\phi)^2\phi^{\prime 2}}{2 \s N^2} 
+ W(\phi) \rp \s d a - \frac{3 \s \pi}{4 \s G} \int_0^\infty \lp N 
+ \frac{1}{N} \rp \s a \s d a
\end{equation}
Variation with respect to $N$ and $\phi$ gives back the system
(\ref{eq:O4:phi},\ref{eq:O4:a}), 
showing that no physical degree of freedom has been lost. 

Since the derivative of $N$ does not appear in Eq.~\eqref{eq:O4:ac3}, 
one can express $N$ as a function of $\phi$ and $\phi'$:
\begin{equation} \label{eq:O4:compN}
N = \lp \frac{1 - 4 \s \pi \s G \s a^2 \s f(\phi)^2\phi^{\prime 2}/3}
{1 - 8 \s \pi \s G \s a^2 \s W(\phi)/3} \rp^{1/2} .
\end{equation}
This quantity is always real. The expression in the denominator is
a recurring and important combination for the eigenvalue problem, 
hence we write
\begin{equation} \label{eq:Qdef}
Q[\phi] \equiv 1 - \frac{8 \s \pi \s G}{3} \s a^2 W(\phi).
\end{equation}
Plugging Eq.~\eqref{eq:O4:compN} into Eq.~\eqref{eq:O4:ac3}, 
we obtain an unconstrained action for the scalar field ${\phi}$,
\begin{equation} \label{eq:O4:ac2}
S = - 
\frac{3 \s \pi}{2 \s G} \int_0^\infty 
\mathrm{sgn}  \lp Q[\phi] \rp 
\left[ Q[\phi] \s
\lp 1 - \frac{4 \s \pi \s G \s a^2}{3} \s f(\phi)^2\phi^{\prime 2} \rp
\right]^{1/2} a \s d a .
\end{equation}
Extremization of this action gives back Eq.~\eqref{eq:O4:phi} with the 
explicit form of $N$ given by Eq.~\eqref{eq:O4:a}. 

This expression for the action can be conveniently used to derive the 
eigenvalue equation.
To this end, let us assume we have an exact solution ${\phi} =\phi_b$. 
We look for a perturbed solution of the form\footnote{Notice that 
$\varphi(a)$ is the geodesic distance, in the metric (\ref{field metric}),
between the perturbed and background fields.}
$\phi = \phi_b +\varphi/f(\phi_b)$. 
To quadratic order in ${\varphi}$, the action reads 
$S = S^{(0)} + S^{(2)} + O \lp{\varphi}^3 \rp$, 
where $S^{(0)}$ is the action of the background instanton and
\begin{equation}
\begin{aligned}
S^{(2)} =  
2 \s \pi^2 \int_0^\infty 
{a^3 \over N_b}\s
\left[ \lp D^2W + \frac{8 \s \pi \s G \s a^2}{3 \s Q_b} \s (DW)^2
+ \frac{8 \s \pi \s G f}{3 \s Q_b } \s \phi_b' \s DW \rp 
\frac{\varphi^2}{2 \s Q_b} + \frac{1}{N_b^2 \s Q_b} \s 
\frac{\varphi^{\prime2}}{2} \right] \s d a .
\label{quadact}
\end{aligned}
\end{equation}
where $Q_b = Q[\phi_b]$, and $D = f^{-1} d/d\phi$.
The simplest way to derive ({\ref{quadact}) is to regard $\phi$ as a 
coordinate on a one dimensional manifold with metric
\begin{equation}
\mathfrak{g}=f(\phi)^2d\phi^2.\label{field metric}
\end{equation}
The action can be evaluated in a coordinate frame with $f=1$, and then 
the general expression is recovered by replacing derivatives with respect to 
$\phi$ by the covariant derivative $D$.

The corresponding eigenvalue equation obtained from the perturbed action is
\begin{equation} \label{eq:O4:eigen}
\frac{1}{N_b\,a^3}
\s \lp \frac{a^3}{N_b^3\,Q_b} \s {\varphi}' \rp' = 
\left[  \frac{1}{N_b^2Q_b} \s \lp D^2W
+ \frac{8 \s \pi \s G \s a^2}{3 \s Q_b} \s \lp DW\rp^2
+ \frac{8 \s \pi \s G \s a^2 f}{3 \s Q_b} \s \phi_b' \s DW \rp
- \lambda \right]{\varphi},
\end{equation}
where $\lambda$ is the eigenvalue. 

By definition, $N_b$ is always positive. However, $Q_b$ will be negative 
wherever $a^2 W(\phi_b) >  3 / (8 \s \pi \s G)$. 
When $Q_b$ is negative, the quadratic action is unbounded from below. 
(In fact, it can reach arbitrarily high negative values 
even for square integrable perturbations of unit $L^2$ norm provided 
the latter oscillate sufficiently fast in the region where $Q_b< 0$.) As was 
shown in~\cite{Lee:2014uza, Koehn:2015hga} for instantons in de Sitter 
space, if the eigenmode equation has no singularity, negativity of the kinetic 
term implies the existence of an infinite number of negative eigenvalues. 

The profusion of negative modes can be qualitatively understood as follows. 
In regions where the kinetic term is positive, for sufficiently large negative 
values of $\lambda$, ${\varphi}$ increases or decreases exponentially with 
$a$, with growth rate $N_b^2\sqrt{Q \lvert\lambda\rvert}$.
If the kinetic term is positive everywhere, the boundary conditions at $a = 0$ 
and $a \to \infty$ can not be simultaneously satisfied. 
If the kinetic term reaches negative values, however, ${\varphi}$ becomes 
oscillatory in some interval, allowing us to match an exponentially decreasing 
function for $a \to \infty$ with a hyperbolic cosine for $a \approx 0$. 
More precisely, they will match provided the difference between the phases 
of the oscillations at both ends of the region where the kinetic term is negative 
exactly compensates the difference between the ratios ${\varphi}' / {\varphi}$ 
for the hyperbolic cosine on the left and the exponential on the right.

It must be noted, however, that these negative modes may be physically 
relevant only for very thin bubbles. Indeed, negativity of the kinetic term requires that 
$\lvert a f(\phi_b) \phi_b' \rvert$ reaches values above the Planck mass. 
In many models, ${\phi}_b$ is limited to be less than $1$ in Planck units, 
so that the semiclassical analysis should not break down. These negative 
modes may thus be physically meaningful only if $| a f(\phi_b) {\phi}_b' | 
\gg |{\phi}_b|$, i.e., either when the width of the bubble is much smaller 
than its radius or when $f$ is large. The latter case can occur when 
$\xi$ is large and negative. In the following section we will see examples 
that realise both of these possibilities.  


\subsection{Numerical results}
\label{sub:O4_num}

\begin{figure}
\includegraphics[width=0.49\linewidth]{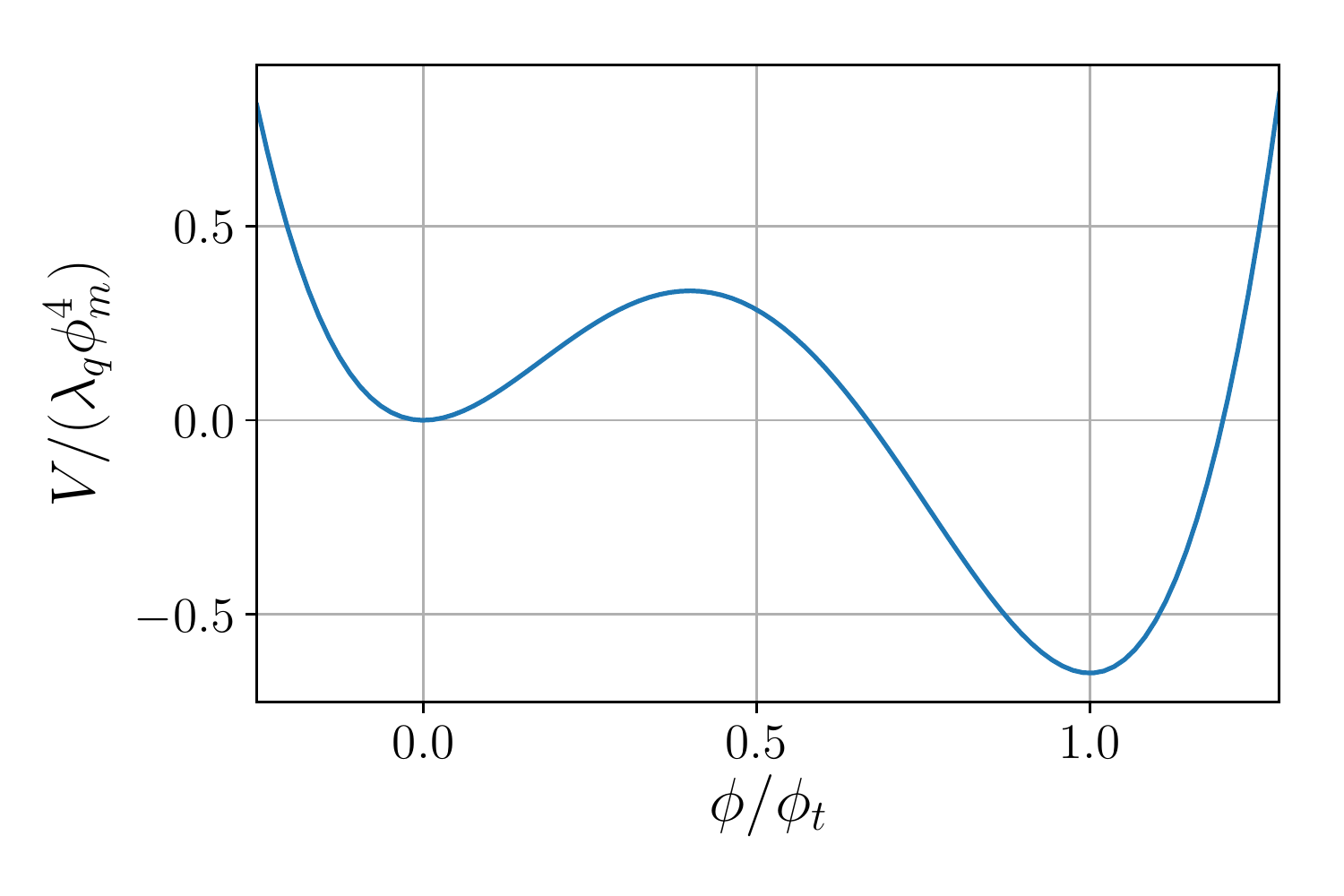}
\includegraphics[width=0.49\linewidth]{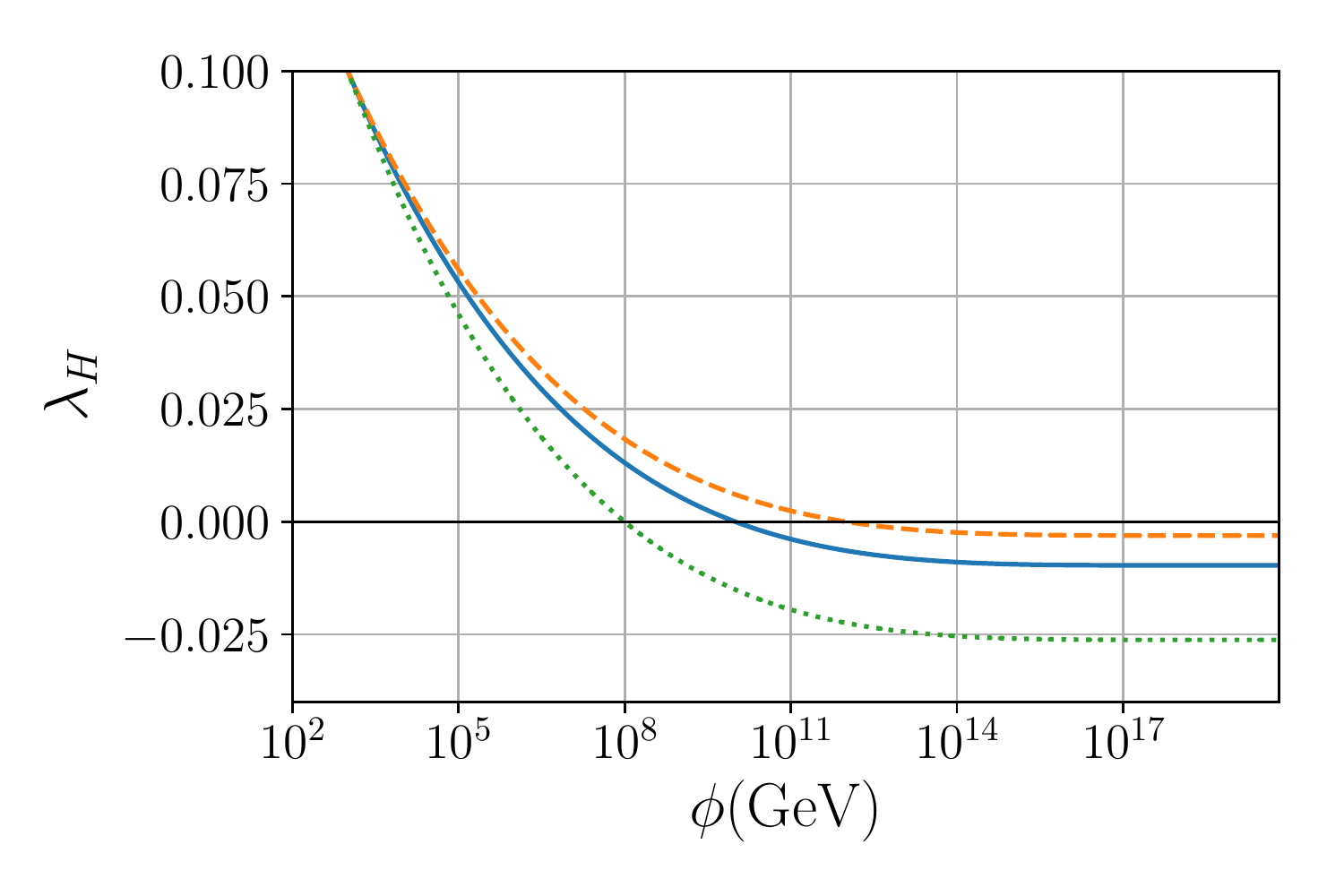}
\caption{
Left panel: Quartic potential~\eqref{eq:V4} for $\phi_m = M_p/10$, $\phi_t = M_p/4$, and $\lambda_q = 10/3$.
Right panel: Effective coupling for the Higgs-like potential~\eqref{eq:Hpot} 
for $\Lambda = 10^8 \mathrm{GeV}$ (green, dotted), 
$\Lambda = 10^{10} \mathrm{GeV}$ (blue, continuous), and 
$\Lambda = 10^{12} \mathrm{GeV}$ (orange, dashed), 
and $q$ chosen so that $\lambda(\phi = 10^3 \mathrm{GeV}) = 0.1$.
} \label{fig:potentials}
\end{figure}
We turn now to the numerical solution of the system~(\ref{eq:O4:phi},\ref{eq:O4:a}) 
and eigenvalue 
equation~\eqref{eq:O4:eigen} with two different shapes for the potential $V$. 
The first case is a quartic potential
\begin{equation}\label{eq:V4}
V_q(\phi) = \frac14 \lambda_q\phi^4-\frac13\lambda_q(\phi_m+\phi_t)\phi^3+
\frac12 \lambda_q\phi_m\phi_t\phi^2,
\end{equation}
which has been parameterised by $\phi_m$ and $\phi_t$, the field values at the 
maximum and the non-zero minimum respectively. 
The parameter $\lambda_q$ sets the overall scale.
The origin $\phi = 0$ is a false vacuum, and $\phi_t$ is the true vacuum
when $\phi_t>2\phi_m$. One example is shown in the left panel of Fig.~\ref{fig:potentials}.
The numerical results do not rely on the thin-wall approximation,
but the latter provides a useful the check on the results. The thin
wall approximation is valid when $\phi_t\sim2\phi_m$.

An important derived  parameter is the AdS radius of the true vacuum $\ell$.
For minimal coupling ($\xi=0$),
\begin{equation}
\ell^2=-{3 M_p^2\over V(\phi_t)}.
\end{equation}
For example, we expect gravitational back-reaction to be important
when the bubble radius is comparable to the AdS radius.
In the thin-wall approximation, the `flat-space' bubble radius $R_0=3\sigma/\epsilon$ 
and the ratio
\begin{equation}
{R_0\over \ell}={1\over\sqrt{2}}{\phi_t\over M_p}
\left(1-2{\phi_m\over\phi_t}\right)^{-1/2}.
\end{equation}
Note that this is independent of the overall scale parameter $\lambda_q$.
It is possible to scan through different values of $R_0/\ell$ by
fixing $\phi_m/\phi_t$ and scanning through different values of $\phi_t$

While the quartic potential is convenient for illustrative purposes, 
obtaining results which may be applicable to the Standard Model 
requires a more realistic one. 
We thus also used a Higgs-like potential of the form 
\begin{equation}\label{eq:Hpot}
V_H(\phi) = \frac{\lambda_H(\phi)}{4} \s \phi^4, \quad \lambda_H(\phi) 
= q \s \lp \lp \ln {\phi\over M_p} \rp^4 - \lp \ln {\Lambda\over M_p} \rp^4 \rp.
\end{equation}
In this expression, $\Lambda > 0$ is the scale at which the coupling
and the potential vanish, and $q$ is a strictly positive number. 
Like the quartic potential $V_q(\phi)$, this potential has a local minimum at $\phi = 0$. 
Plots of the function $\lambda_H$ for three different choices of $(q, \Lambda)$ 
are shown in the right panel of Fig.~\ref{fig:potentials}. 
They approximate the next-to-next-to-leading order calculations reported 
in~\cite{Degrassi:2012ry} with different values of the top quark mass.

The height of the Higgs potential barrier is small compared to $\Lambda^4$, making
the bubble solutions shallow, with thick walls, and Higgs values inside the bubble 
extending beyond the barrier but do not reaching a true vacuum. The potential
inside the bubble is roughly of order $\Lambda^4$ and the bubble
size is of order $\Lambda^{-1}$, so that the `effective' value of $R_0/\ell$
in this case is around $\Lambda/M_p$.

\begin{figure} 
\includegraphics[width=0.49\linewidth]{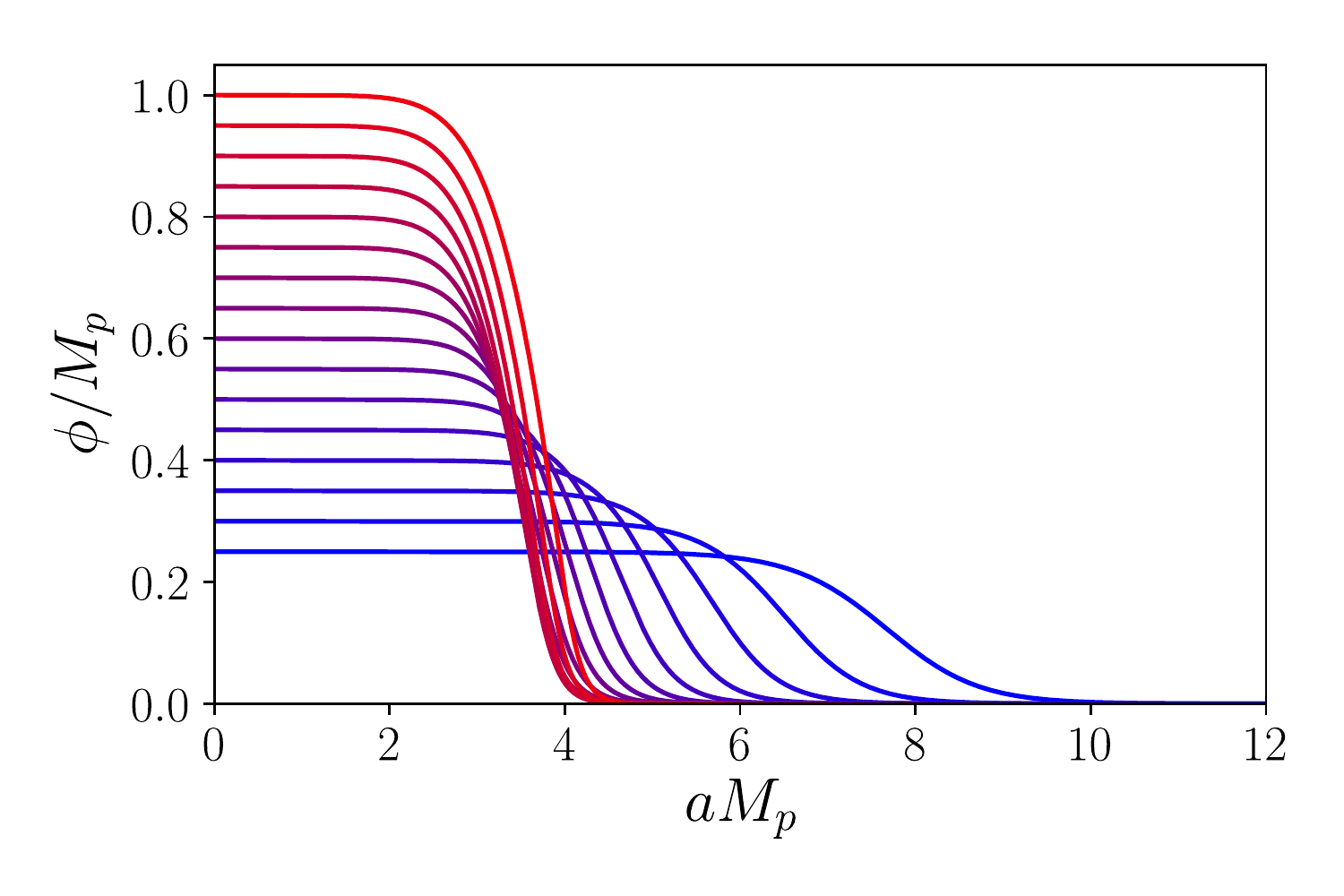}
\includegraphics[width=0.49\linewidth]{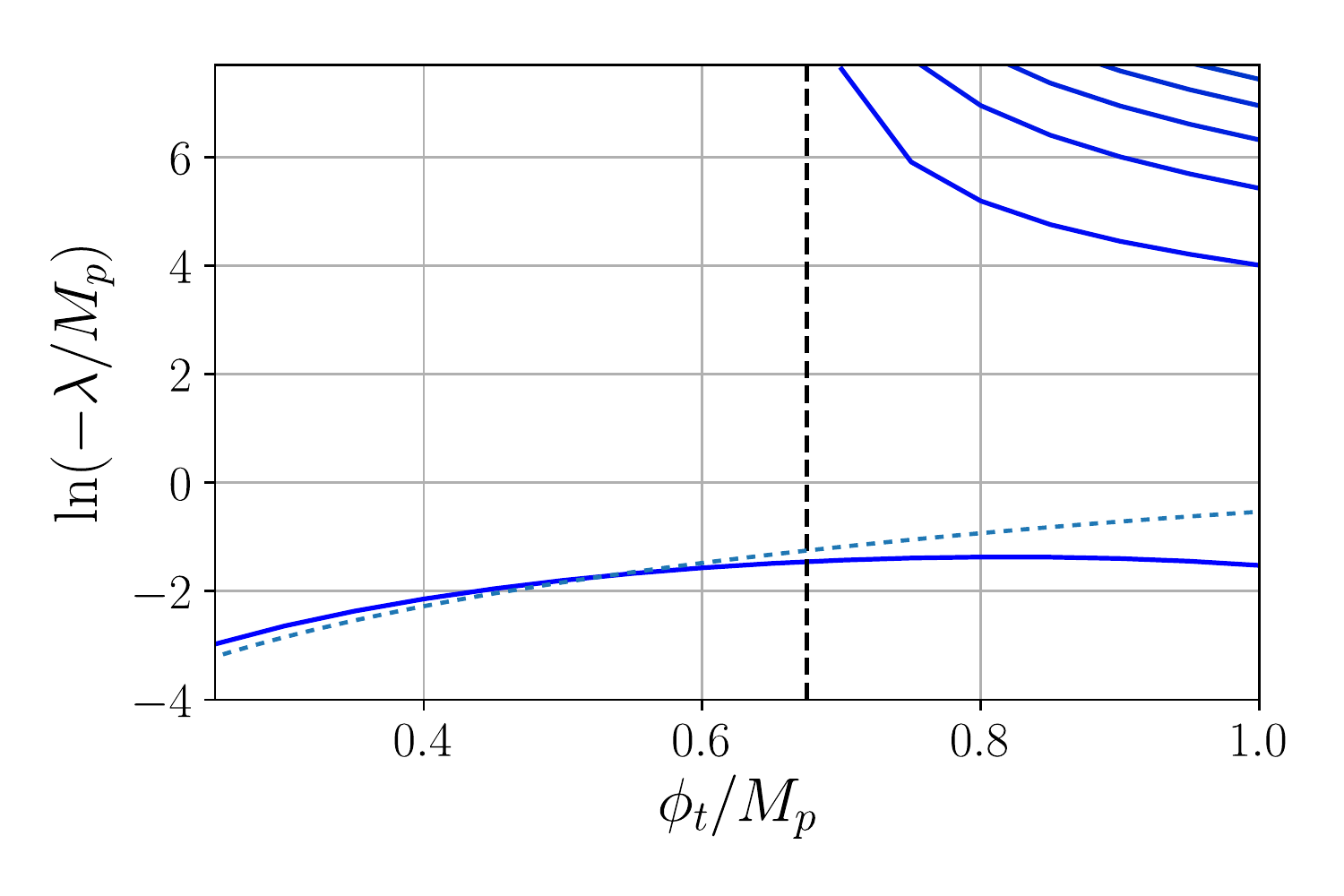}
\caption{Left panel: $\mathrm{O}(4)$-symmetric instantons obtained 
with the quartic potential~\eqref{eq:V4} for 
$\lambda_q = 128$ and $\phi_t / \phi_m = 2.5$.
The value of $\phi_t$ in Planck units increases from blue to red. 
Right panel: Negative eigenvalues for these solutions. The dashed curve shows
the thin wall approximation (\ref{evapprox}).
The vertical dashed line shows the value $\phi_{c}$ of $\phi_t$ above which $Q_b$  
takes negative values.}\label{fig:O4_ev} 
\end{figure}
We first work with the quadratic potential and $\xi = 0$, i.e., with a minimal coupling 
between the scalar  field $\phi$ and gravity. In Fig.~\ref{fig:O4_ev} we show the negative 
eigenvalues with fixed ratio $\phi_t / \phi_m = 2.5$, 
$\lambda_q = 128$, and $\phi_t$ ranging from $0.25M_p$ to $M_p$.
Below a critical value $\phi_{c}$, here close to $0.67M_p$, there is only one negative mode.
The dashed line shows the negative mode obtained for the thin-wall approximation 
using (\ref{evapprox}), which agrees quite well with the numerical result 
despite the walls not being particularly thin.
 
The quantity $Q$ defined in (\ref{eq:Qdef}) is positive for the bubble solutions with 
$\phi_t<\phi_{c}$, but for $\phi_t > \phi_{c}$, $Q$ takes negative values in a 
finite interval of $a$. 
Correspondingly, we find new negative eigenvalues, all but one going to $- \infty$ in the 
limit $\phi_t \to \phi_{c}$, in agreement with our approximate analysis
in Eq.~\eqref{eq:O4:estnegla}. The numerical evidence therefore supports the
existence of infinitely many negative eigenvalues for $\phi_t > \phi_{c}$.

Results with nonminimal coupling are shown in Fig.~\ref{fig:O4_q_ev}.
\begin{figure}
\includegraphics[width=0.49\linewidth]{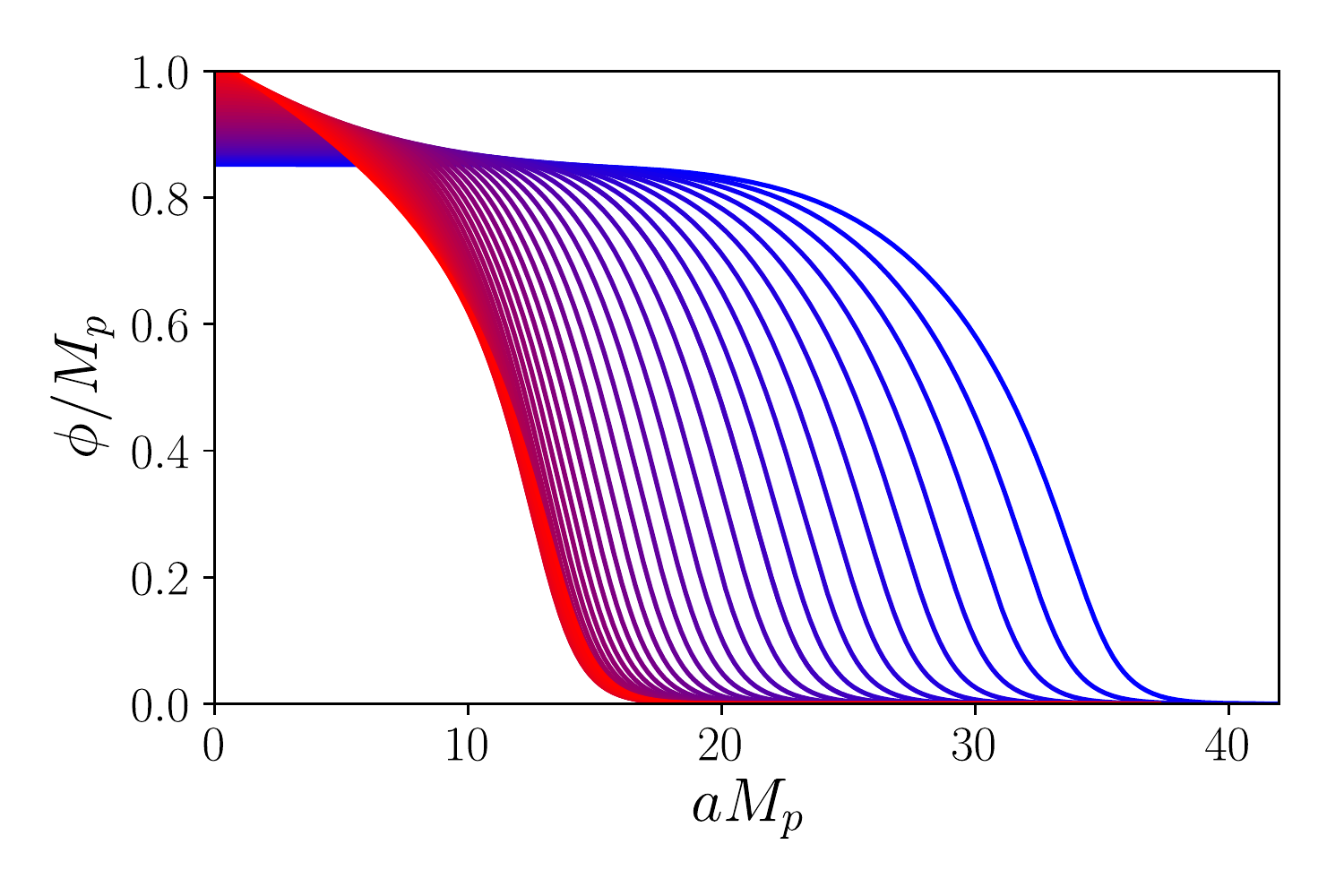}
\includegraphics[width=0.49\linewidth]{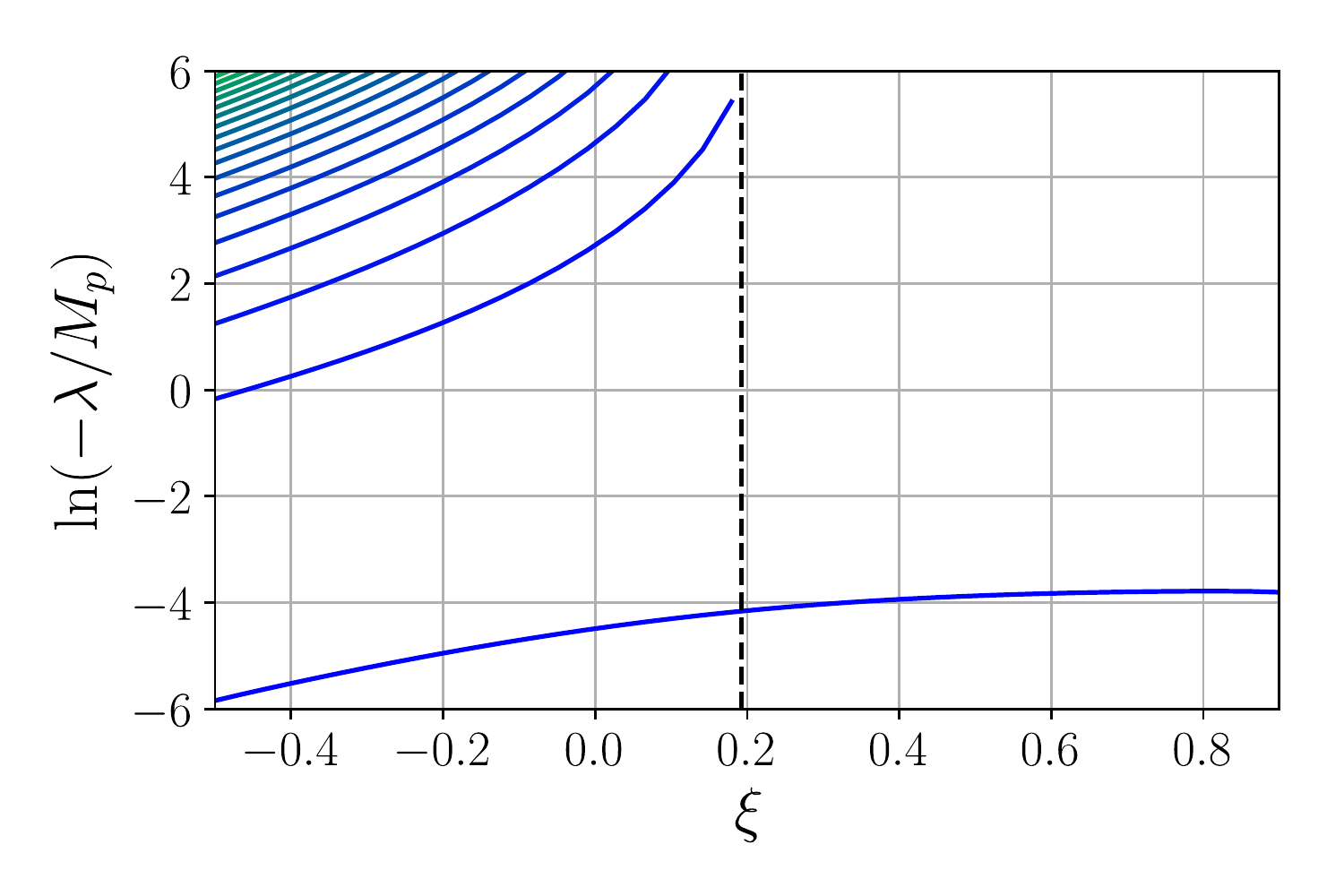}
\caption{Left panel: $\mathrm{O}(4)$-symmetric instantons obtained with the 
quartic potential~\eqref{eq:V4} for $\phi_m = 0.36$, $\phi_t = 0.84$, and $\lambda_q = 10/3$, for different 
values of the nonminimal coupling $\xi$ ranging from $-0.5$ to $0.9$. 
The value of $\xi$ increases from blue to red. 
Right panel: Negative eigenvalues for these solutions. 
The vertical dashed line shows the value $\xi_c$ of $\xi$ below which $Q_b$ 
takes negative values. 
} \label{fig:O4_q_ev}
\end{figure}
Here the parameters of the potential are fixed to $\phi_m = 0.36$, $\phi_t = 0.84$,  
and $\lambda_4 = 10/3$, and the nonminimal coupling $\xi$ is varied between 
$-0.5$ and $0.9$. At the level of the instanton solution, the main effect of a 
negative value of $\xi$ seems to be to increase the radius of the bubble, 
while a positive value increases $\phi(0)$. Its role is more dramatic when 
considering the negative modes: as shown in the right panel of the 
figure, there is a critical value $\xi_c$, here close to $0.2$, above which only 
one negative mode is present, but below which there is an infinite number of them. 
As already noticed when varying $\phi_t$, the first case corresponds to a 
positive $Q$, while in the second case this function takes negative values 
in a finite interval of $a$. As in the previous case also, all but one negative 
eigenvalues go to negative infinity when approaching the threshold $\xi \to \xi_c$. 

Figures~\ref{fig:O4_H_em},~\ref{fig:O4_H_ev}, and~\ref{fig:O4_xi_ev} 
show results obtained with the potential $V_H$. 
To ease the numerical resolution, they are made with relatively high values 
of $\Lambda$, close to unity in Planck units. 
We found a similar behavior for smaller values of this parameter. 
\begin{figure}
\centering
\includegraphics[width=0.49\linewidth]{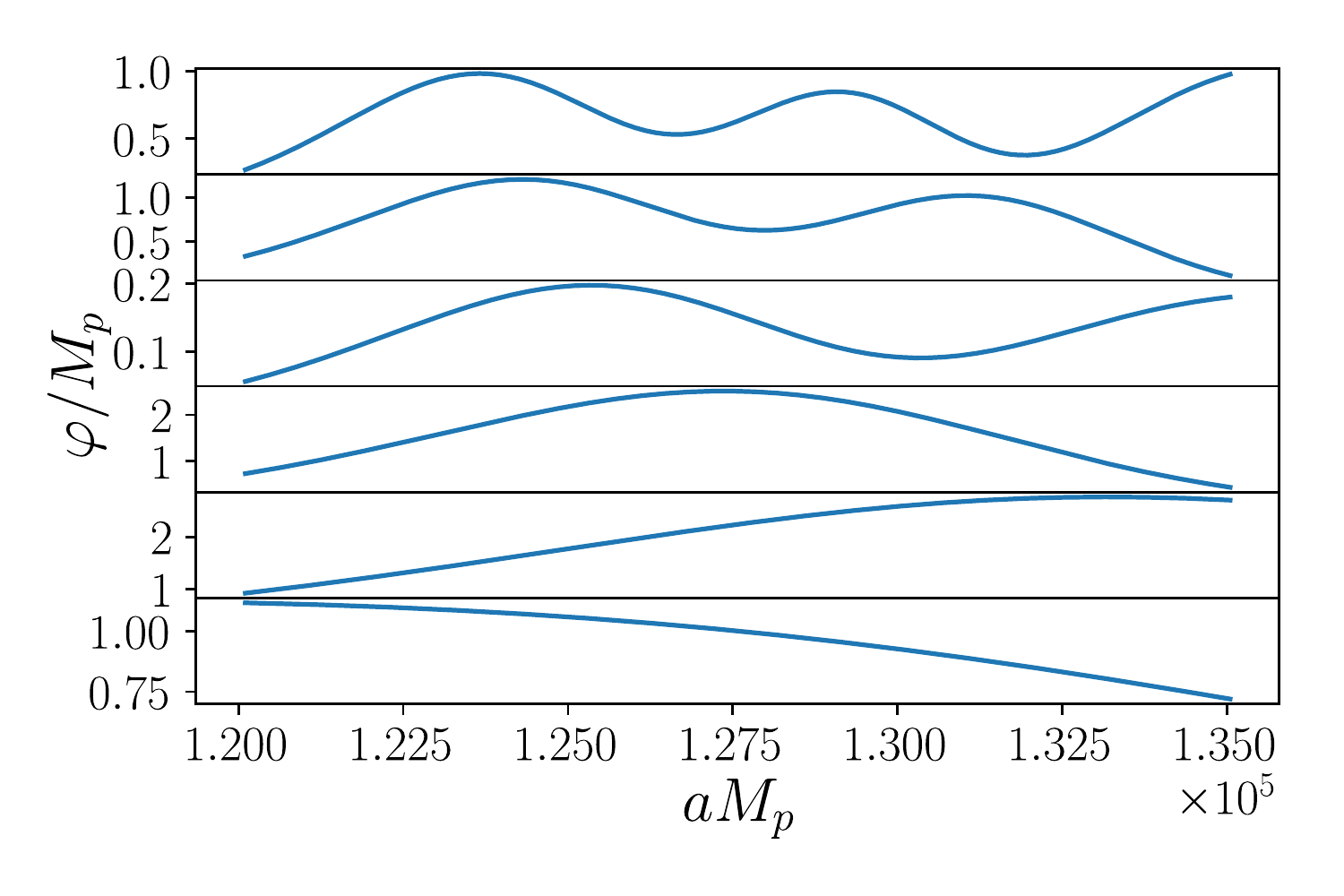}
\caption{Plots of the first six negative modes in the region where the kinetic 
term is negative. We use the Higgs-like potential~\eqref{eq:Hpot} with 
$q = 10^{-7}$ and $\Lambda = 0.3$, and a minimal coupling $\xi = -5.3$, 
slightly below the critical one $\xi_c \approx -4.8$ for this potential. 
(The normalization is arbitrary.)
} \label{fig:O4_H_em}
\end{figure}
In Fig.~\ref{fig:O4_H_em} are shown the first six negative modes for fixed 
potential and a minimal coupling $\xi$ slightly smaller than $\xi_c$, in the 
region where $Q< 0$. The main information is that, as expected, negative 
modes are oscillatory in this region, and that the $n^\textrm{th}$ one 
has approximately $n/2$ wavelengths for sufficiently large $n$.
\begin{figure}
\includegraphics[width=0.49\linewidth]{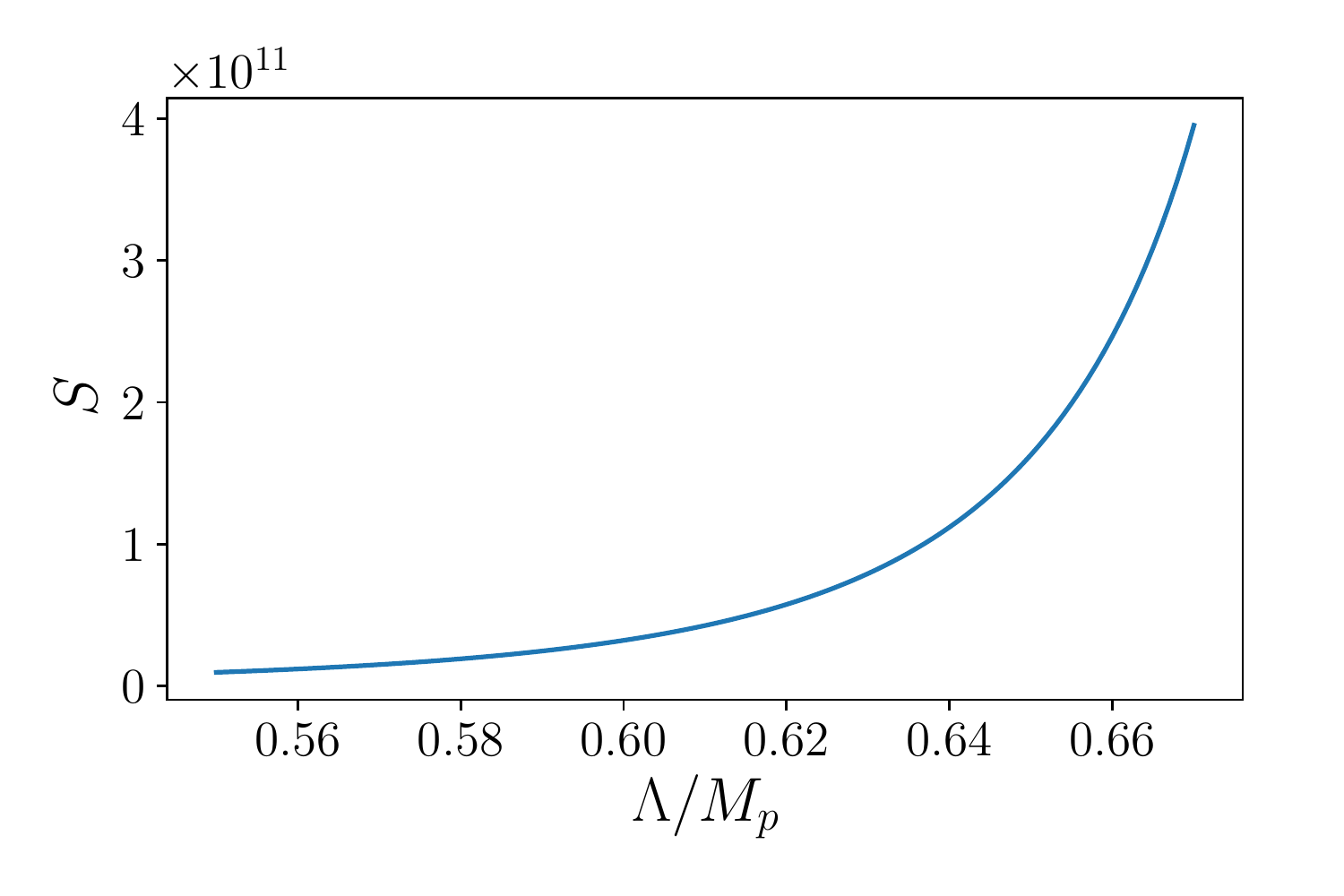}
\includegraphics[width=0.49\linewidth]{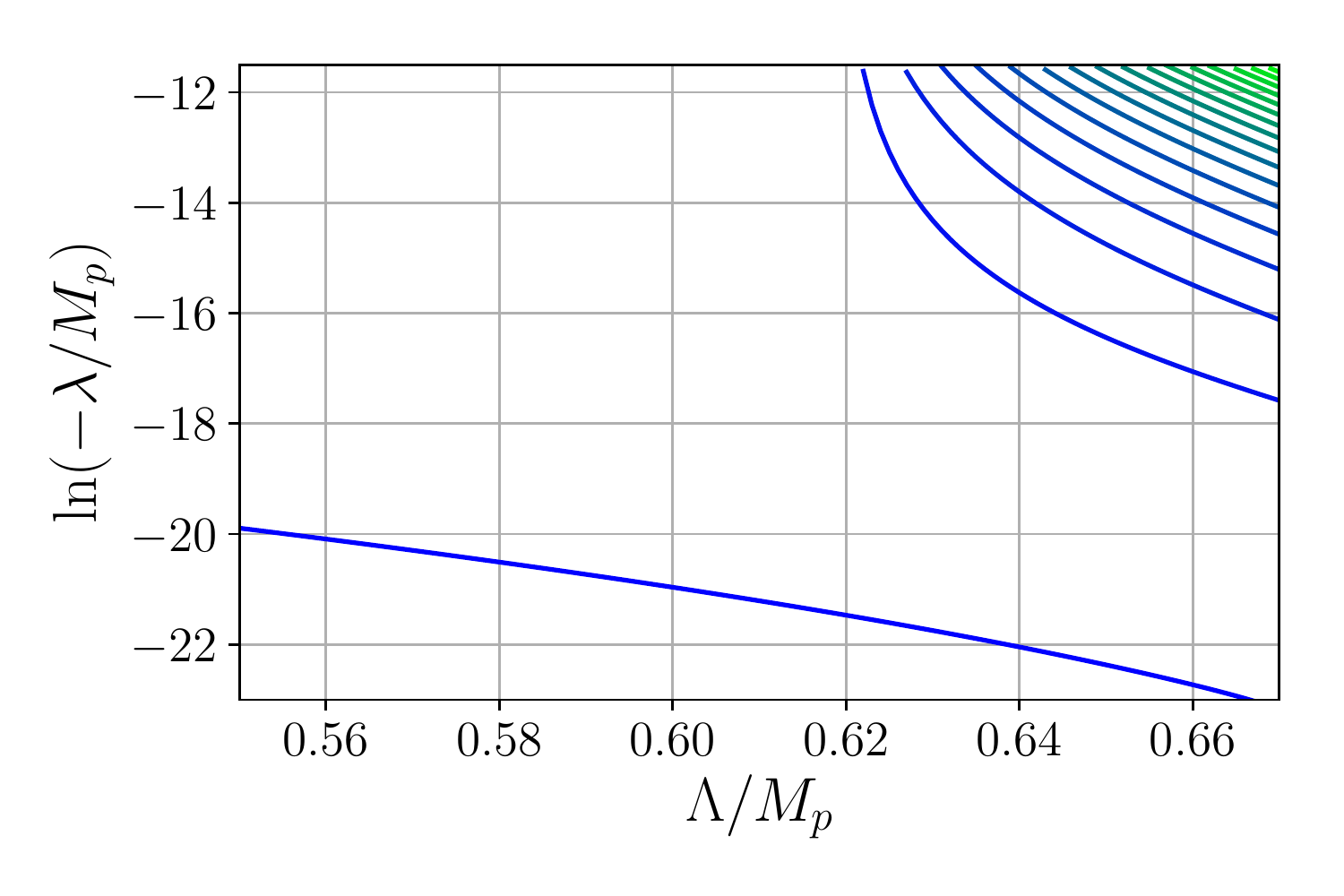}
\caption{Euclidean action (left panel) and negative eigenvalues (right panel) 
of an asymptotically flat $\mathrm{O}(4)$-symmetric instanton with the 
Higgs potential~\eqref{eq:Hpot} with $q = 10^{-7}$ and $\xi=0$
.} \label{fig:O4_H_ev}
\end{figure}

\begin{figure}
\includegraphics[width=0.49\linewidth]{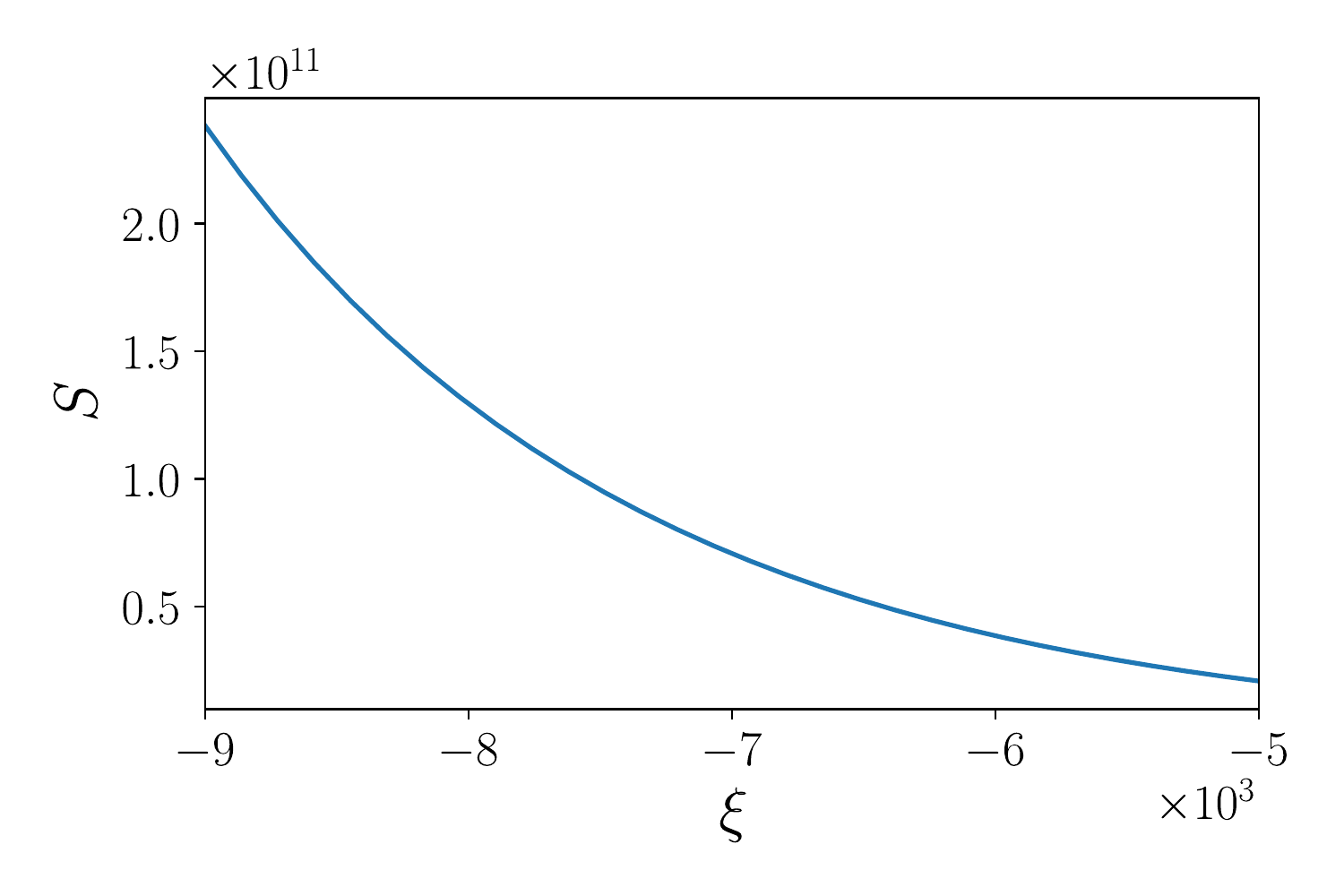}
\includegraphics[width=0.49\linewidth]{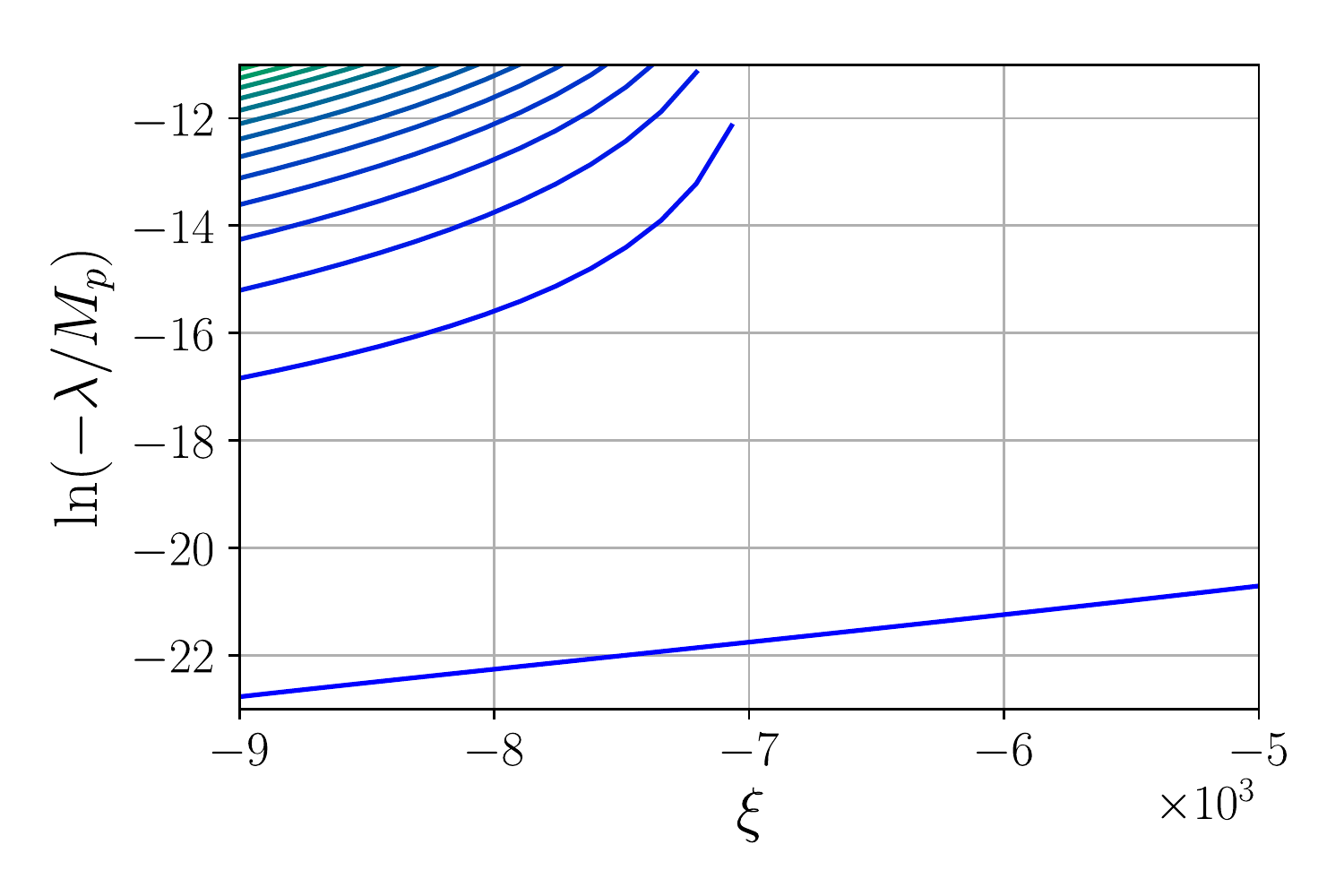}
\caption{Euclidean action (left panel) and negative eigenvalues (right panel) 
of an asymptotically flat $\mathrm{O}(4)$-symmetric instanton for the 
Higgs potential~\eqref{eq:Hpot} with $q = 10^{-7}$ and $\Lambda = 0.5$
.} \label{fig:O4_xi_ev}
\end{figure}
Figures ~\ref{fig:O4_H_ev} and~\ref{fig:O4_xi_ev} shows the Euclidean 
action and negative eigenvalues of instantons as functions of $\Lambda$ 
and $\xi$ respectively, for $q = 10^{-7}$.  As can be seen on the left panels 
and more generally in Figure~\ref{fig:xic}, the Euclidean action of instantons supporting 
infinitely many negative modes is huge, making the transition rate negligible. 
We found the same holds for all parameters we tried. 
It thus seems that, for realistic potentials, the appearance of an infinite 
number of negative eigenvalues requires such a strong back-reaction from 
gravity on the instanton that the probability of bubble nucleation becomes 
negligibly small. Conversely, all instantons we found which gave non-negligible 
decay rates have only one negative eigenvalue. 

\begin{figure}
\includegraphics[width=0.49\linewidth]{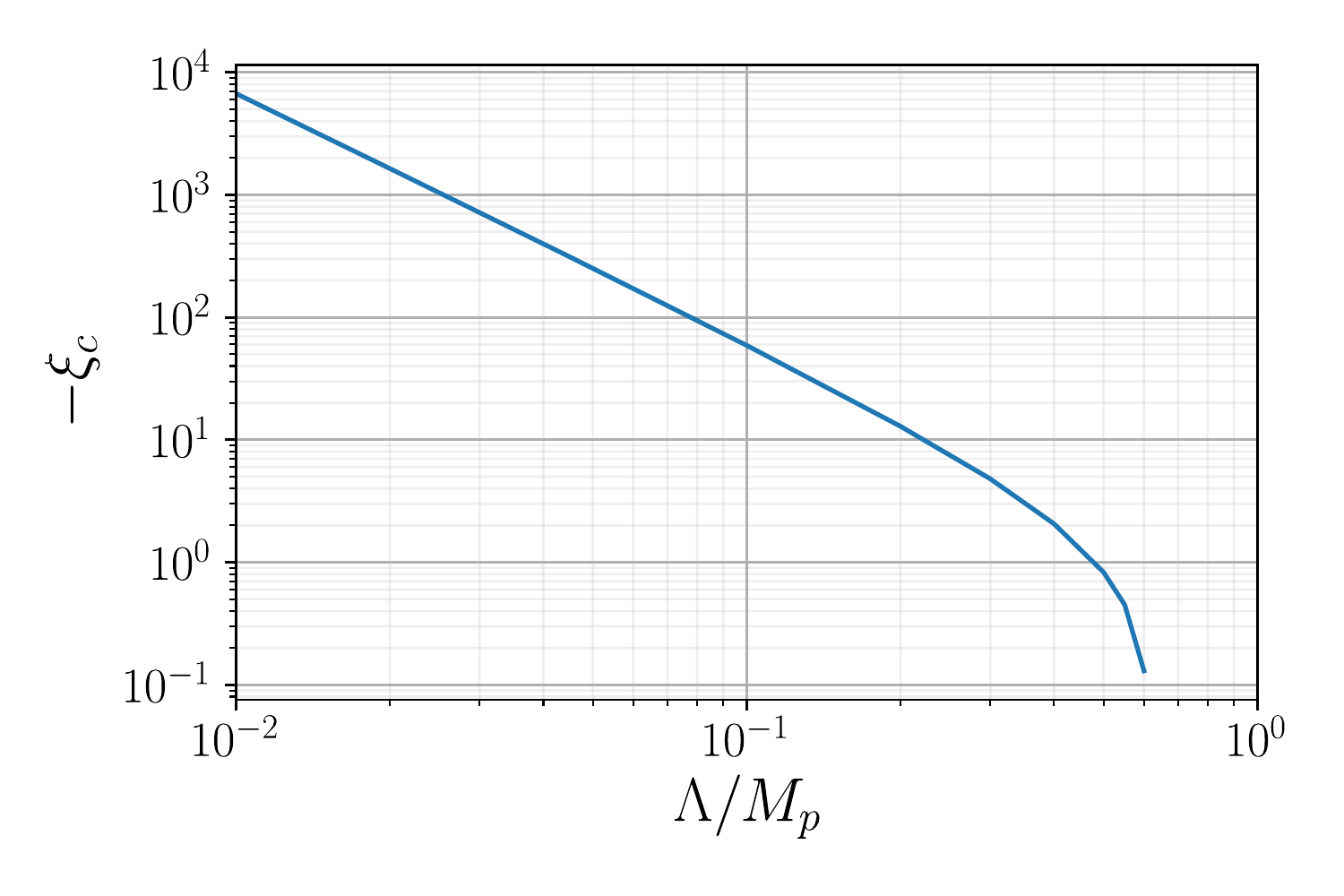}
\includegraphics[width=0.49\linewidth]{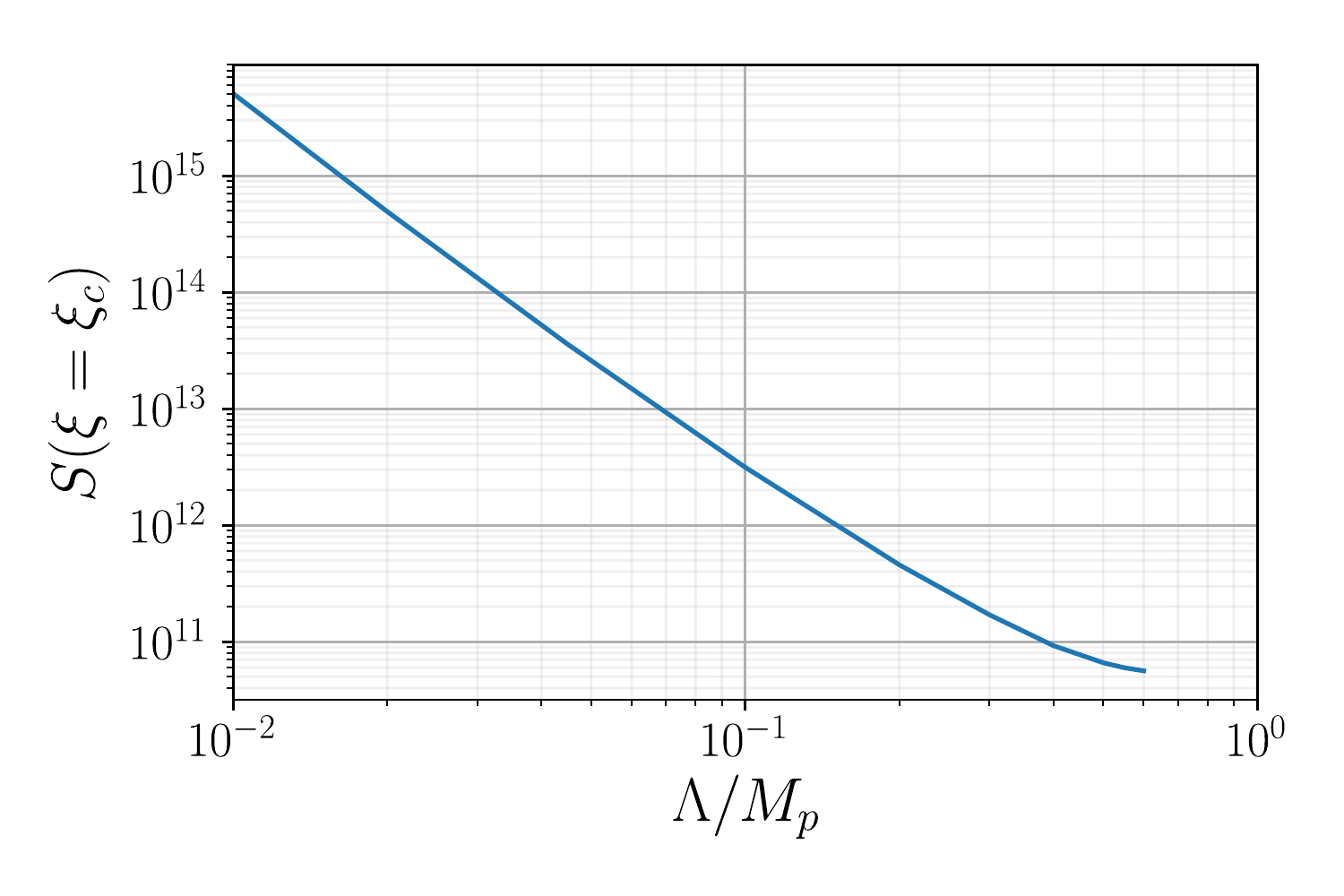}
\caption{Left panel: Dependence of the critical value $\xi_c$ of the 
nonminimal coupling below which an infinite number of negative modes 
is present in the scale $\Lambda$ at which the Higgs potential vanishes. 
The potential is given by~\eqref{eq:Hpot} with $q = 10^{-7}$. 
For larger values of $\Lambda$, $\xi_c$ is formally positive, but $\phi$ 
reaches values close to the Planck scale so that the semi-classical 
approximation is not expected to be valid.  
Right panel: Euclidean action of the critical instanton with $\xi = \xi_c$ 
for the same values of $\Lambda$.}\label{fig:xic}
\end{figure}

\subsection{Analytical estimates} 
\label{sec:anest}

We now mention two analytical results which help understand the 
numerical observations reported above. 
We first give an estimate of the large negative eigenvalues. 
For large values of $-\lambda$, one can neglect the other terms in the 
right-hand side of Eq.~\eqref{eq:O4:eigen}. 
Moreover, since the rate of change of ${\varphi}$ is proportional to 
$\sqrt{|\lambda|}$, we can in this limit neglect the variations of $a$, $Q$, 
and $N_b$. The eigenvalue equation thus becomes
\begin{equation} 
{\varphi}'' \approx - Q \s N_b^2 \s \lambda \s {\varphi}.
\end{equation}
Let us call $a_-$ and $a_+$, the boundaries of the interval in which 
$Q$ is negative, ordered as $a_- < a_+$. Then, ${\varphi}$ is exponentially 
increasing or decreasing for $a > a_+$, and oscillating for $a_- < a < a_+$. 
The global solution will be decreasing at infinity provided the oscillating 
solution for $a$ just below $a_+$ can be matched with the decaying one 
for $a > a_+$, with one or the opposite sign. This occurs twice each time 
we add one wavelength in the interval $[a_-, a_+]$. One thus expects that, 
for large values of $n$, the $n^\text{th}$ negative eigenvalue $\lambda_n$ satisfies
\begin{equation}
\int_{a_-}^{a_+} \sqrt{QN_b^2 \s \lambda_n} \s d a \approx \pi \s n,
\end{equation}
i.e.,
\begin{equation} \label{eq:O4:estnegla}
\lambda_n \approx \frac{- \pi^2 \s n^2}{\lp \int_{a_-}^{a_+} \sqrt{- Q} \s N_b \s d a \rp^2}.
\end{equation}

Notice that, since the $n^\text{th}$ negative mode oscillates with a wave 
vector proportional to $\sqrt{- \lambda_n}$ in the region where $B^{(0)}$ 
is negative, it must have an amplitude proportional to $(-\lambda)^{-1/4}$ 
to be normalized for the Klein-Gordon inner product. 
From the above estimate, $(-\lambda)^{-1/4} \propto n^{-1/2}$. 
One can thus expect that the sum of the contributions from negative modes 
to quadratic observables are formally divergent, which may point to an 
instability of the background solution or, as conjectured in
\cite{Lavrelashvili:1985vn}, to a breakdown of the semiclassical approximation. 
However, a more detailed analysis, would be required to ascertain 
this\footnote{In particular, logarithmic corrections to Eq.~\eqref{eq:O4:estnegla}, 
if present, could make the series convergent.}.

Next, we estimate the critical value $\xi_c$ below which an infinite number 
of negative modes are present. We assume that $- \xi \gg 1$ and that 
$- \xi \s \phi \gg 1 / \sqrt{G}$ in the relevant domain of $a$ 
(typically inside the bubble and including a significant fraction of the wall).
In this limit, we can integrate to find the canonical field $\tilde\phi$, defined by
\begin{equation}
\tilde\phi=\int d\phi f(\phi)\approx\sqrt{\frac{3}{16 \s \pi \s G}} \ln \lp 1 
- 8 \s \pi \s G \s \xi \s \phi^2 \rp.
\label{eq:O4:tildephivsphi_approx}
\end{equation}
The modified potential~\eqref{eq:O4:W} thus becomes
\begin{equation}
W(\phi) \approx 
\e^{- 8 \tilde\phi\sqrt{\pi G/3}} \,\,V \lp 
\sqrt{1 - \e^{4 \tilde\phi\sqrt{\pi G/3}} \over 8 \s \pi \s G \s \xi} \rp .
\end{equation}
Let us assume that $V$ has a zero at a value $\Lambda > 0$ of $\phi$. 
We call $\tilde{\Lambda}$ the corresponding value of ${\tilde\phi}$. 
If the potential has no other typical scale, the maximum value reached by 
$a \s \tilde\phi'$ should be of order $\tilde{\Lambda}$. 
Denoting by $\alpha$ the ratio $a\tilde\phi'/\tilde\Lambda$, we have
\begin{equation}
\min Q = 1 - \frac{4 \s \pi \s G}{3} \s \alpha^2 \s \tilde{\Lambda}^2.
\end{equation}
Using Eq.~\eqref{eq:O4:tildephivsphi_approx}, this may be rewritten as
\begin{equation}
\min Q \approx 1 - \frac{\alpha^2}{4} \s \ln \lp 1 - 8 \s \pi \s G \s \xi \s \Lambda^2 \rp^2.
\end{equation}
This quantity is negative provided $\xi < \xi_c$, where 
\begin{equation}
\xi_c \approx \frac{1 - \e^{2 / \alpha}}{8 \s \pi \s G \s \Lambda^2}. 
\end{equation}
We thus expect $- 8 \s \pi \s G \s \Lambda^2 \s \xi_c$ to be of order $1$ 
for sufficiently small values of $\Lambda^2 \s G$, in accordance with results 
shown in the left panel of Figure~\ref{fig:xic}.  

\section{Vacuum decay with a black hole}

Now we turn to vacuum decay seeded by a microscopic black hole. 
The initial state consists of a region of false vacuum containing a 
Schwarzschild black hole. In Euclidean space, the black
hole metric is periodic in the imaginary time coordinate with period 
$\beta=8\pi GM$. We enforce the same boundary conditions on the 
bubble solution to ensure that the tunnelling exponent 
$B=S_E[\phi_b]-S_E[\phi_{\rm fv}]$ is finite. The black hole inside the 
bubble solution has a smaller mass than the original black hole. 

When we take static (i.e.\ independent of Euclidean time $\tau$) 
solutions, there is a remarkable simplification in the expression for 
the action which allows to to express the tunnelling exponent in terms of 
the reduction in black hole entropy \cite{Burda:2016mou},
\begin{equation}
B={{\cal A}_S\over 4G}-{{\cal A}_R\over 4G},\label{exponent}
\end{equation}
where ${\cal A}_S$ and ${\cal A}_R$ are the areas of the event horizon 
of the black hole seed and the black hole remnant. The bubble solution 
has a conical singularity at the horizon, but when this is properly taken into 
account there is no ambiguity in the action \cite{Gregory:2013hja}. 

Note that in general one can find instanton solutions with a range of
remnant mass for a given seed mass, but there is a unique remnant mass
with lowest action. There are then two branches of solutions~\cite{Gregory:2013hja}:
One branch comprises non-static instantons that are a variant of the CDL 
instanton, and continuously connected to this $O(4)$ symmetric solution
in the limit $M \to 0$. The other branch occurs for seed masses larger than
some critical mass, $M_C$, and is a `static' solution. These solutions are 
relevant for black holes above the Planck mass, where one can trust the 
semi-classical methods used.
As shown in~\cite{Gregory:2013hja}, the static instanton is the relevant
instanton for Higgs vacuum decay, thus in this section we consider static 
instantons only. These have the further advantage that they are dependent 
only on the radial coordinate. Since the static branch is not continuously 
connected to the CDL instanton, we do not expect to recover the results of 
Section~\ref{sec:O4} in the limit $M \to 0$.

\subsection{Model and field equations}

We consider the real scalar field $\phi$ minimally coupled to gravity 
with the Einstein-scalar action (\ref{eq:gaction}). We look for 
spherically-symmetric bubble solutions where $\phi$ depends only on a 
radial coordinate $r$ and the metric has the form: 
\begin{equation}
ds^2 = f(r) \s \e^{2 \s \delta(r)} \s d \tau^2 + \frac{d r^2}{f(r)} + r^2 d \Omega_2^2,
\end{equation}
where $\tau$ is the Euclidean time, $f$ is a smooth positive function, 
and $d \Omega_2^2$ is the metric on a unit-radius, two-dimensional sphere. 
It is also convenient to define the function $\mu$ by 
\begin{equation}
f(r) = 1 - \frac{2 \s G \s \mu(r)}{r}. 
\end{equation}
The Einstein equations then give~\cite{Burda:2016mou}
\begin{align} 
\label{eq:BH_FEf}
& \lp r^2 \s \e^{\delta} \s f \s \phi' \rp'
= r^2 \s \e^{\delta} \s V'[\phi], \\
\label{eq:BH_FEm}
& \mu' = 4 \s \pi \s r^2 \s \lp \frac12 f \s \phi^{\prime2} + V[\phi] \rp, \\
\label{eq:BH_FEd}
& \delta' = 4 \s \pi \s G\, r \s \phi^{\prime2} .
\end{align}
We look for asymptotically flat black hole solutions, for which $f(r)$ vanishes 
at the horizon $r=r_h$ and 
$\phi$ approaches the false vacuum as $r\to\infty$.
Without loss of generality (up to a global rescaling of $\tau$), one can 
impose $\delta(r_h) = 0$. 
The final boundary condition is given by a regularity condition at the 
horizon~\cite{Burda:2016mou}:
\begin{equation}
\phi'(r_h) = \frac{r_h \s V'[\phi(r_h)]}{1 - 8 \s \pi \s G \s r_h^2 \s V[\phi(r_h)]}. 
\end{equation}

In order to compute the Euclidean action, we require the Ricci scalar,
\begin{equation}
R = - f'' - 3 \s \delta' \s f' - 2 \s \delta'' \s f - 2 \delta^{\prime2} \s f 
- \frac{4}{r} \s f' - \frac{4}{r} \s \delta' \s f + \frac{4 \s G}{r^3} \s \mu.
\end{equation}
Using this and performing an integration by parts, the Einstein-Hilbert action~\eqref{eq:gaction} becomes
\begin{equation}\label{eq:BH:EHaction}
S_\mathrm{E}[\phi_b] =
4 \s \pi \s \beta \int_{r_h}^\infty \e^{\delta} \s r^2 \s \lp 
- \frac{\mu'}{4 \s \pi \s r^2}
+ \frac{f\phi^{\prime2}}{2} + V \rp d r
+ \frac{\beta}{2} \s \lp M_S \s \e^{\delta(\infty)} - M_R + r_h \s \mu'(r_h) \rp,
\end{equation}
where $\beta$ denotes the period in $\tau$, $M_S \equiv \mu(\infty)$ is the 
ADM mass of the initial black hole, 
and $M_R \equiv \mu(r_h)$ is the mass of the residual black hole. 
The first term vanishes when imposing Eq.~\eqref{eq:BH_FEm}.
In Refs \cite{Gregory:2013hja} and~\cite{Burda:2015yfa}, it was shown 
that including boundary terms $S_{\partial}$ gives the result
\begin{equation}
S_E[\phi_b] + S_{\partial}[\phi_b] = - \frac{\mathcal{A}_R}{4 \s G}+\beta M_S.
\end{equation}
The false vacuum black hole gives
\begin{equation}
S_E[\phi_{\rm fvbh}] +  S_{\partial}[\phi_{\rm fvbh}] 
= - \frac{\mathcal{A}_S}{4 \s G}+\beta M_S.
\end{equation}
The difference produces the tunnelling exponent Eq.\ (\ref{exponent}).

To determine the eigenvalue equation, we write $\phi = \phi_b + \phi_1$, 
$\mu = \mu_b + \mu_1$, 
and $\delta = \delta_b + \delta_1$, where $(\phi_b, \mu_b, \delta_b)$ is an 
exact solution of Eqs.~(\ref{eq:BH_FEf} -- \ref{eq:BH_FEd}).
We define $f_b \equiv 1 - 2 \s G \s \mu_b / r$.
To quadratic order, and discarding a boundary term, the action reads 
$S = S^{(0)} + S^{(2)} + \dots$, where $S^{(0)}$
is the action of the instanton, dots represent higher-order terms, and
\begin{equation}
\begin{aligned}
S^{(2)} = {} & 
4 \s \pi \s \beta \int_{r_h}^\infty r^2 \s \e^{\delta_b } \s
\left[ f_b \s \phi_b' \s \phi_1' - \frac{G\mu_1}{r} \s \phi_b^{\prime2} 
+ V' [ \phi_b ] \phi_1 - \frac{\mu_1'}{4 \s \pi \s r^2} \right] d r \\
& + 4 \s \pi \s \beta \int_{r_h}^\infty r^2 \s \e^{\delta_b } \s \delta_1  \s 
\left[ f_b \s \phi_b' \s \phi_1' - \frac{G\mu_1}{r} \s \phi_b^{\prime2} 
+ V' [\phi_b] \phi_1 - \frac{\mu_1'}{4 \s \pi \s r^2} \right] d r \\
& + 4 \s \pi \s \beta \int_{r_h}^\infty r^2 \s \e^{\delta_b} \s
\left[ \frac{f_b}{2} \s \phi_1^{\prime2} 
- \frac{2 \s G\mu_1}{r}  \s \phi_b' \s \phi_1'
+ V'' [ \phi_b ]\frac{\phi_1^2}{2} \right] d r.
\end{aligned}
\end{equation}
The first integral vanishes with the boundary condition $\mu_1(\infty) = 0$. 
Variation of $S^{(2)}$ with respect to $\mu_1$ gives the constraint
\begin{equation} \label{eq:BW:const_delta1}
\delta_1' = 8 \s \pi \s G \s r \s \phi_b' \s \phi_1'. 
\end{equation} 
Variation with respect to $\delta_1$ gives
\begin{equation}
\mu_1' = 4 \s \pi \s r^2 \s \left[ f_b \s \phi_b' \s \phi_1' 
+ V' [ \phi_b ]\phi_1
- \frac{G\mu_1}{r} \s \phi_b^{\prime2}
\right].
\end{equation}
Using equations~(\ref{eq:BH_FEf} -- \ref{eq:BH_FEd}) and assuming the 
boundary condition $\mu_1(\infty) = \phi_1(\infty) = 0$, this becomes
\begin{equation} \label{eq:BW:const_mu1}
\mu_1 = 4 \s \pi \s \s r^2 \s f_b \s \phi_b' \s \phi_1. 
\end{equation}
Using Eqs.~\eqref{eq:BW:const_delta1} and~\eqref{eq:BW:const_mu1}, 
the quadratic action becomes
\begin{equation}
S^{(2)} = 
4 \s \pi \s \beta \int_{r_h}^\infty r^2 \s \e^{\delta_b} \s
\left[ \frac{f_b}{2} \s \phi_1^{\prime2}
- 8 \s \pi \s G \s r \s f_b \s \phi_b^{\prime2}\s \phi_1 \s \phi_1'
+ V'' [ \phi_b ] \frac{\phi_1^2}{2} \right] d r.
\end{equation}
Integrating by parts the second term inside the square brackets and using again 
Eqs.~(\ref{eq:BH_FEf} -- \ref{eq:BH_FEd}), one obtains
\begin{equation} \label{eq:BH_s2_s}
S^{(2)} = 4 \s \pi \s \beta \int_{r_h}^\infty r^2 \s \e^{\delta_b } \s
\left[ \frac{f_b}{2} \s \phi_1^{\prime2}
+ \mathcal{V}(r) \s \frac{\phi_1^2}{2} \right] d r,
\end{equation}
where 
\begin{equation}
\mathcal{V}(r) \equiv V'' [\phi_b ]+ 16 \s \pi \s G \s r \s V'(\phi_b) \s \phi_b'
- 8 \s \pi \s G \s r \s \lp f_b' + \delta_b' \s f_b + \frac{f_b}{r} 
\rp \phi_b'^2.
\end{equation}
The eigenvalue equation from the action~\eqref{eq:BH_s2_s} is:
\begin{equation} \label{eq:BW_eigen}
\frac{e^{-\delta_b }}{r^2} \s
\frac{d}{d r} \lp r^2 \s e^{\delta_b } \s f_b \s \phi_1' \rp = 
\lp \mathcal{V}(r) - \lambda \rp \s \phi_1.
\end{equation}

Contrary to the $\mathrm{O}(4)$-symmetric case studied in 
Section~\ref{sec:O4}, here the prefactor of the 
kinetic term in the quadratic action, $r^2 \s \e^{\delta_b } \s f_b $, 
is always positive outside the horizon. 
Since the presence of an infinite number of negative modes in the 
previous case was due to the kinetic term reaching negative 
values, we conjecture they do not arise in the present model\footnote{As 
explained in Section~\ref{sec:O4}, a nonminimal 
coupling to gravity is equivalent to a change of potential. In the present case, 
this will not change the sign of the kinetic term.}.
This conjecture is supported by the numerical investigation of 
Eq.~\eqref{eq:BW_eigen} discussed below. 

\subsection{Numerical results}

We solved the system~(\ref{eq:BH_FEf} -- \ref{eq:BH_FEd}) and the 
eigenvalue equation~\eqref{eq:BW_eigen} in 
the two potentials~\eqref{eq:V4} and~\eqref{eq:Hpot}. 
Results for the tunneling exponent $B$ and negative eigenvalues $\lambda$ 
are shown in Figs.~\ref{fig:BH_B_la} and~\ref{fig:BH_q_B_la}. 

Notice that in the case of the quartic potential~\eqref{eq:V4} we have an 
approximate symmetry when the effects of gravity are sufficiently small. 
Indeed, neglecting the term in $\mu'$, equation~\eqref{eq:BH:EHaction} is 
invariant under $\phi \to \eta \s \phi$, 
$\phi_m \to \eta \s \phi_m$, $\phi_t \to \eta \s \phi_t$, $r \to r / \eta$, 
$\beta \to \beta / \eta$, $\mu \to \eta \s \mu$ at 
fixed $a_4$ for any $\eta > 0$. 
The differences between the curves shown in each panel of 
Fig.~\ref{fig:BH_B_la} are thus entirely due to the 
gravitational back-reaction, which has the tendency to increase the 
tunneling exponent $B$ and decrease the absolute value of $\lambda$. 

For both potentials, in the whole range of parameters we tried we always 
found only one negative mode, as could be expected from the facts that the 
kinetic term in the eigenvalue equation~\eqref{eq:BW_eigen} is positive definite 
outside the horizon and the background solution has no node.  
This is the main result of our work, and suggests that the static instantons with 
black holes found in~\cite{Gregory:2013hja,Burda:2015yfa,Burda:2016mou} 
can be safely interpreted as the dominant 
contribution to the decay rate of the false vacuum in the presence of small black holes. 

To confirm and better understand this result, it is useful to define the coordinate 
$x$ by $d x = r^2 \s \e^{\delta_b} \s d r$. 
The eigenvalue equation~\eqref{eq:BW_eigen} then becomes
\begin{equation} 
\frac{d}{d x} \lp r^4 \s \e^{2 \s \delta_b} \s f_b \s \frac{d \phi_1}{d x} \rp = 
\lp \mathcal{V} - \lambda \rp \s \phi_1.
\end{equation}
This has the form of a Schr\"odinger equation, for which nodal theorems apply. 
In particular, the results of~\cite{1995JMP....36.4553A} 
(see also~\cite{doi:10.1137/1.9780898719222}) 
motivate that the number of negative eigenvalues is equal to the number of 
nodes of the solution with $\lambda = 0$ 
satisfying the correct boundary condition at the horizon\footnote{This does 
not constitute a rigorous proof, however,  for two reasons. First, the 
function $f_b$ vanishes at $r=r_h$, while the theorem proved 
in~\cite{1995JMP....36.4553A} applies to uniformly positive functions. 
Second, the boundary condition at the origin used in this reference is 
$\phi_1(0) = 0$ instead of $\phi_1'(r_h) = 0$. We expect that these two 
differences do not change the result, but have so far not been able to 
prove it rigorously.}.
Three solutions corresponding to different values of $r_h$ are shown 
in Fig.~\ref{fig:BH_zeroNmodes} for the 
Higgs-like potential with $\Lambda = 10^{-10}$. 
Each of them has only one node, which confirms there should exist one 
and only one negative mode over each instanton.
\begin{figure}
\centering 
\includegraphics[width=0.49\linewidth]{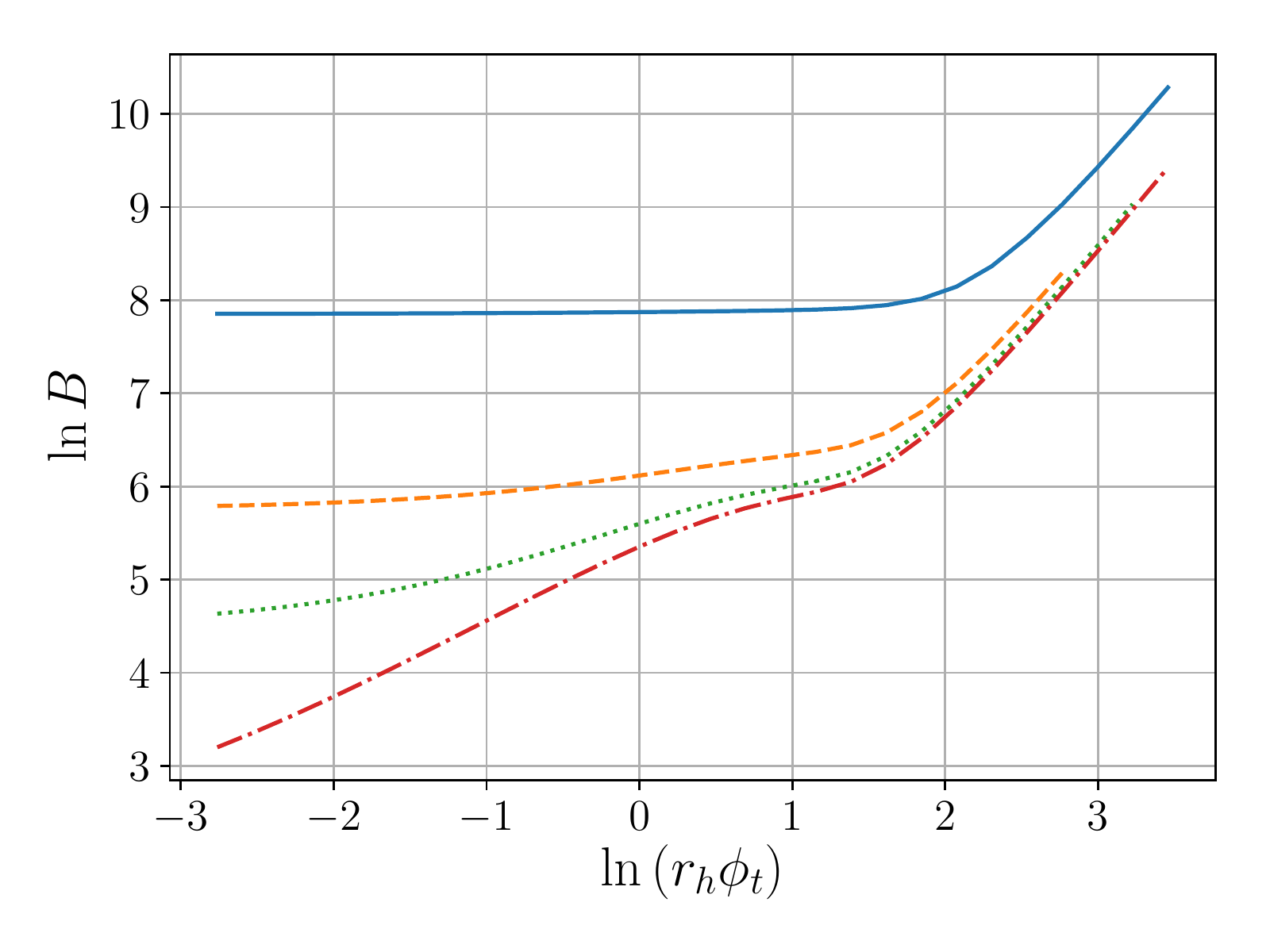}
\includegraphics[width=0.49\linewidth]{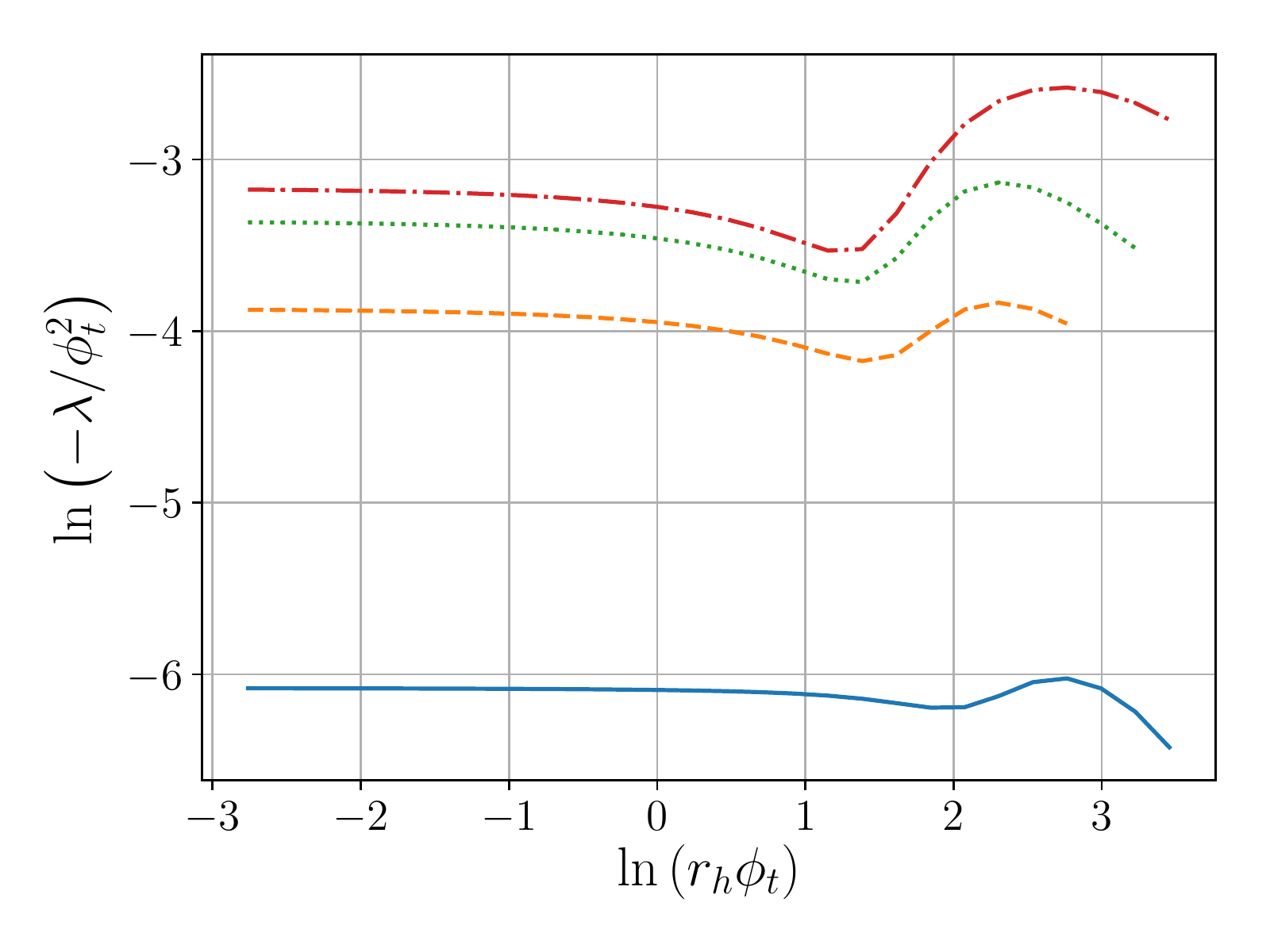}
\caption{
Tunnelling exponent~\eqref{exponent} for seeded nucleation (left panel) and 
negative eigenvalue (right panel) of the instanton with black hole for the 
quartic potential~\eqref{eq:V4} with the parameters $a_4 = 1$, $\phi_t = 2 
\s \alpha$ and $\phi_m = 0.6 \s \alpha$, where $\alpha = 1$ (blue), $10^{-1/4}$ 
(orange), $10^{-1/2}$ (green), and $10^{-1}$ (red). 
}
\label{fig:BH_B_la}
\end{figure}

\begin{figure}
\centering 
\includegraphics[width=0.49\linewidth]{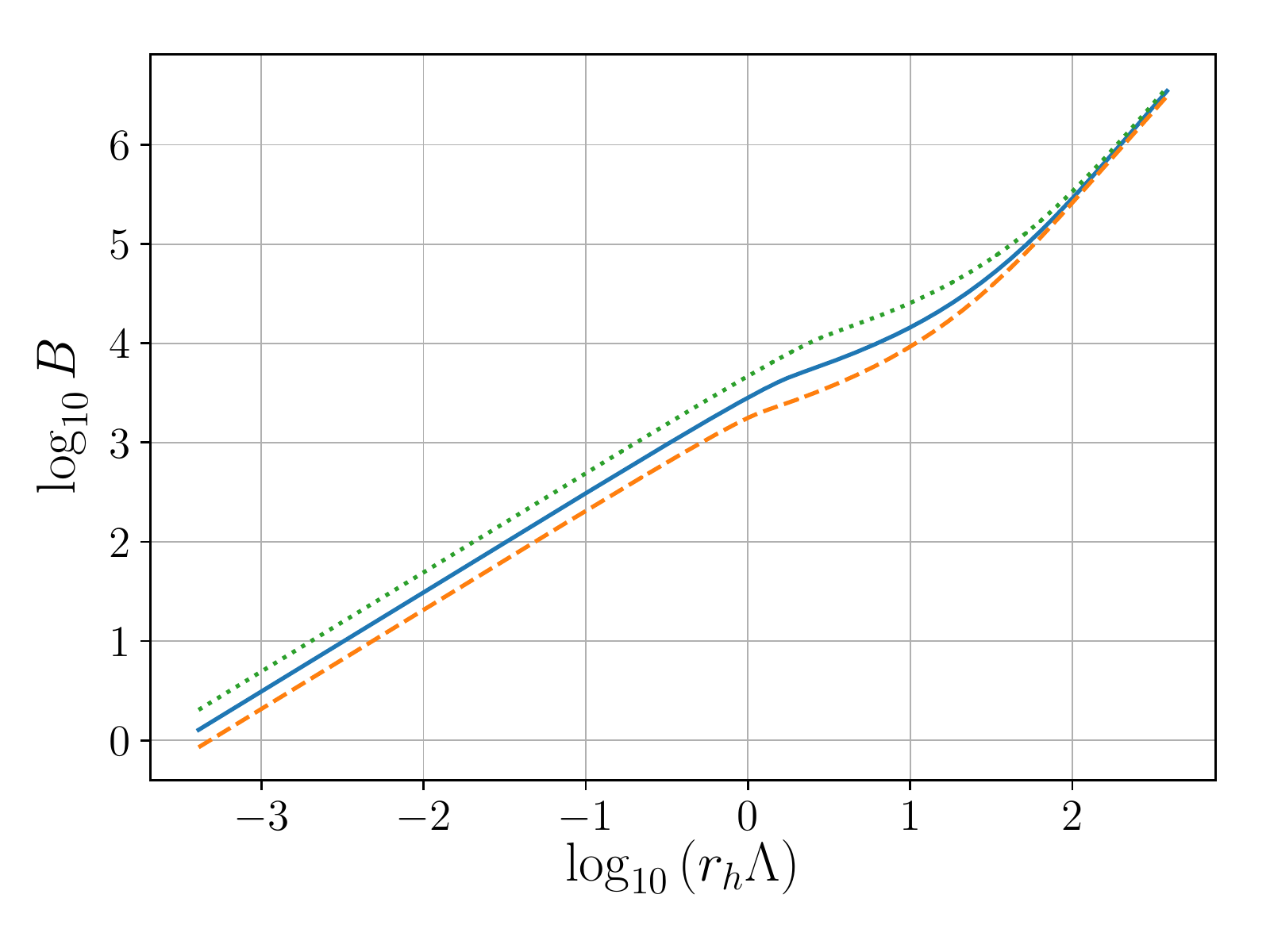}
\includegraphics[width=0.49\linewidth]{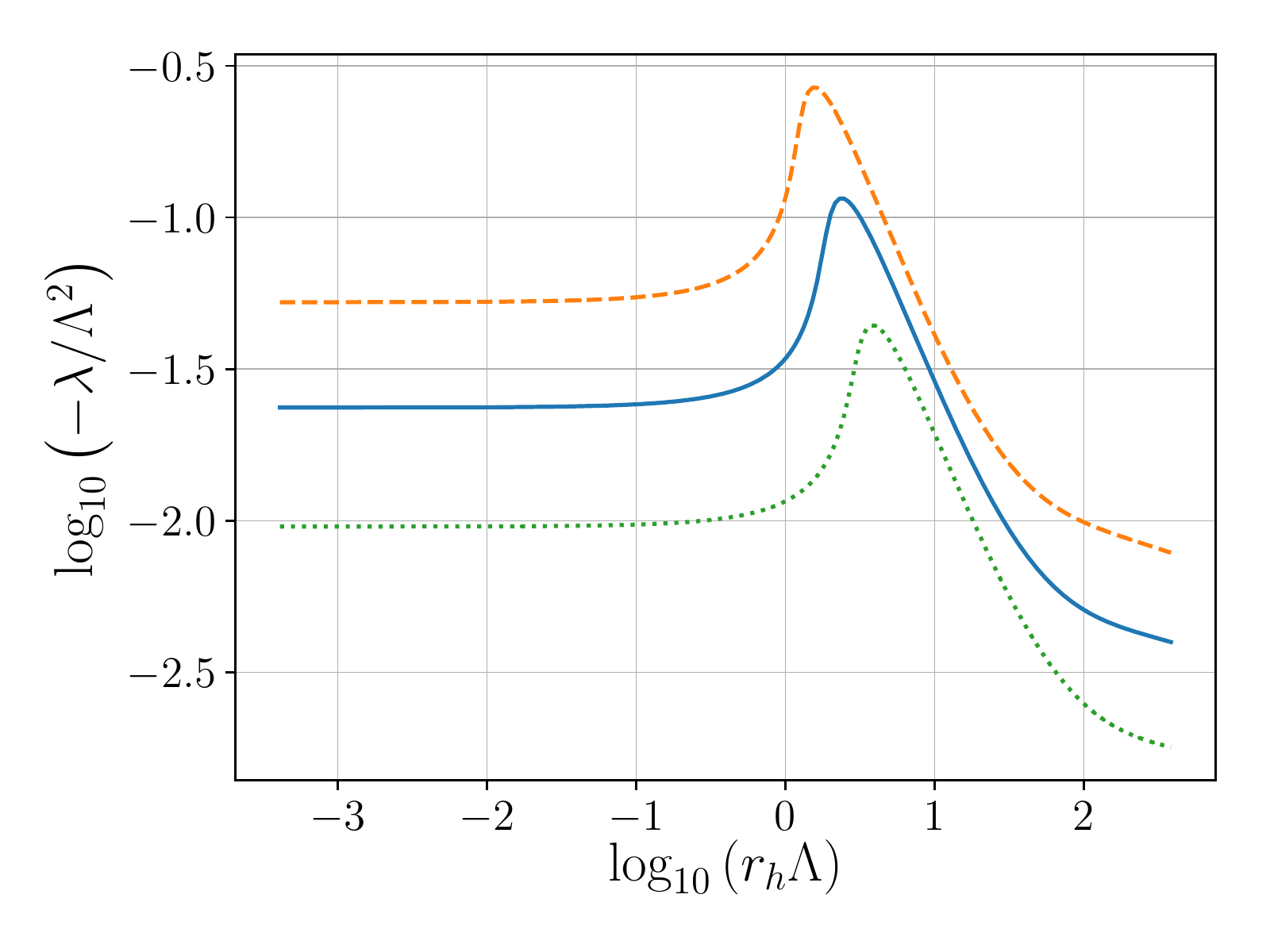}
\caption{
Tunnelling exponent~\eqref{exponent} for seeded nucleation (left panel) 
and negative eigenvalue (right panel) obtained for the Higgs 
potential~\eqref{eq:Hpot} for the same values of the parameters as 
in Fig.~\ref{fig:potentials}, right panel.}\label{fig:BH_q_B_la}
\end{figure}

\begin{figure}
\centering 
\includegraphics[width=0.49\linewidth]{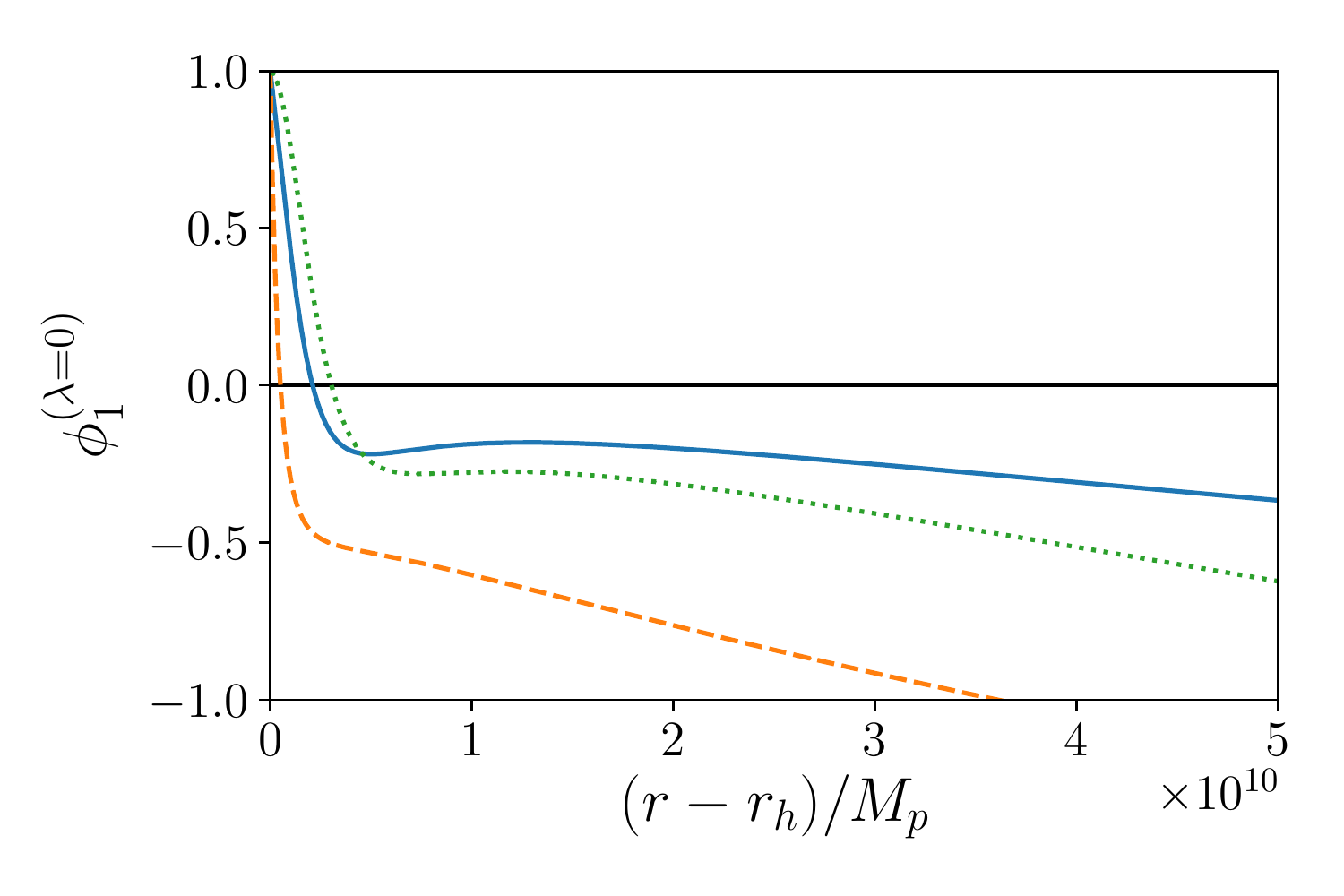}
\caption{Solutions of Eq.~\eqref{exponent} with $\lambda = 0$ for the 
Higgs-like potential~\eqref{eq:Hpot} with $\Lambda = 10^{-10}$. 
The Schwarzschild radius $r_h$ is equal to $0.1 \s \Lambda^{-1}$ (orange), 
$\Lambda^{-1}$ (blue), and $10 \s \Lambda^{-1}$ (green).} 
\label{fig:BH_zeroNmodes}
\end{figure}

\section{Conclusions} 

We have studied negative modes of instantons in two different setups: First 
the case of asymptotically flat $\mathrm{O}(4)$-symmetric ``Coleman-de Luccia'' 
type instantons, including a non-minimal coupling of the scalar, and Second
the case of black hole catalysed vacuum decay developed in
\cite{Burda:2015isa,Burda:2015yfa,Burda:2016mou}.  

For the $\mathrm{O}(4)$-symmetric asymptotically flat instantons, we explored
a wide range of parameter space with a conventional quartic potential, as well
more phenomenologically realistic analytic fit to the Standard Model Higgs
potential. For any value of the non-mimimal coupling parameter $\xi$,
it is always possible to find a region of parameter space in the potential that
has an infinite tower of negative modes for the corresponding instanton, 
however, these parameter values correspond to energies close to the
Planck scale. 

For the black hole instantons, the kinetic term of the quadratic action 
is always positive outside the horizon, and we confirmed numerically
that there is always only one negative mode. 
Although we did not explicitly consider a nonminimal coupling here, 
this would amount to a change of potential which does not affect 
the sign of the kinetic term. 
We thus expect the number of negative modes to be still equal to $1$ 
when including it.

As already noted in~\cite{Lee:2014uza, Koehn:2015hga, Battarra:2012vu}, 
the infinite tower of negative modes arising when the kinetic 
term of the quadratic action reaches negative values remains 
mysterious, although it is intriguing that the tower of modes appear
approximately at the self-compactification scale corresponding to
a domain wall topological defect of tension $\sigma$ 
\cite{Gibbons:1993in,Bonjour:1999kz}.
In section~\ref{sec:anest}, using an analytical estimate for the large 
negative eigenvalues in the $\mathrm{O}(4)$-symmetric case, we argue 
that these infinite negative modes induce a divergence in quadratic 
observables, which seems to support the argument that they may 
signal a breakdown of the semiclassical approximation. 

A more precise answer may require studying time-dependent perturbations 
to see if these additional modes manifest themselves, maybe 
as an instability of the instanton. 
However, assuming asymptotic flatness, both in the $\mathrm{O}(4)$-symmetric 
and black-hole case, we found that realistic instantons 
always have exactly one negative mode. 
It thus seems safe to interpret the lowest-action instanton as giving the 
leading contribution to the tunnelling rate. 

\acknowledgments

This research was supported in part by the Leverhulme Trust, 
in part by Perimeter Institute for Theoretical Physics,
and in part by STFC. KM, FM, and IGM would like to thank Perimeter Institute
for hospitality while this work was in progress. RG, FM and IGM are
supported by Leverhulme grant RPG-2016-233, and
KM is supported by an STFC studentship.
Research at Perimeter Institute is supported by the Government of Canada 
through the Department of Innovation, Science and Economic Development 
and by the Province of Ontario through the Ministry of Research and Innovation.

\bibliography{biblio}

\begin{thebibliography}{48}%
\makeatletter
\providecommand \@ifxundefined [1]{%
 \@ifx{#1\undefined}
}%
\providecommand \@ifnum [1]{%
 \ifnum #1\expandafter \@firstoftwo
 \else \expandafter \@secondoftwo
 \fi
}%
\providecommand \@ifx [1]{%
 \ifx #1\expandafter \@firstoftwo
 \else \expandafter \@secondoftwo
 \fi
}%
\providecommand \natexlab [1]{#1}%
\providecommand \enquote  [1]{``#1''}%
\providecommand \bibnamefont  [1]{#1}%
\providecommand \bibfnamefont [1]{#1}%
\providecommand \citenamefont [1]{#1}%
\providecommand \href@noop [0]{\@secondoftwo}%
\providecommand \href [0]{\begingroup \@sanitize@url \@href}%
\providecommand \@href[1]{\@@startlink{#1}\@@href}%
\providecommand \@@href[1]{\endgroup#1\@@endlink}%
\providecommand \@sanitize@url [0]{\catcode `\\12\catcode `\$12\catcode
  `\&12\catcode `\#12\catcode `\^12\catcode `\_12\catcode `\%12\relax}%
\providecommand \@@startlink[1]{}%
\providecommand \@@endlink[0]{}%
\providecommand \url  [0]{\begingroup\@sanitize@url \@url }%
\providecommand \@url [1]{\endgroup\@href {#1}{\urlprefix }}%
\providecommand \urlprefix  [0]{URL }%
\providecommand \Eprint [0]{\href }%
\providecommand \doibase [0]{http://dx.doi.org/}%
\providecommand \selectlanguage [0]{\@gobble}%
\providecommand \bibinfo  [0]{\@secondoftwo}%
\providecommand \bibfield  [0]{\@secondoftwo}%
\providecommand \translation [1]{[#1]}%
\providecommand \BibitemOpen [0]{}%
\providecommand \bibitemStop [0]{}%
\providecommand \bibitemNoStop [0]{.\EOS\space}%
\providecommand \EOS [0]{\spacefactor3000\relax}%
\providecommand \BibitemShut  [1]{\csname bibitem#1\endcsname}%
\let\auto@bib@innerbib\@empty
\bibitem [{\citenamefont {Coleman}(1977)}]{Coleman:1977}%
  \BibitemOpen
  \bibfield  {author} {\bibinfo {author} {\bibfnamefont {Sidney~R.}\
  \bibnamefont {Coleman}},\ }\bibfield  {title} {\enquote {\bibinfo {title}
  {{The Fate of the False Vacuum. 1. Semiclassical Theory}},}\ }\href {\doibase
  10.1103/PhysRevD.15.2929, 10.1103/PhysRevD.16.1248} {\bibfield  {journal}
  {\bibinfo  {journal} {Phys. Rev.}\ }\textbf {\bibinfo {volume} {D15}},\
  \bibinfo {pages} {2929--2936} (\bibinfo {year} {1977})},\ \bibinfo {note}
  {[Erratum: Phys. Rev.D16,1248(1977)]}\BibitemShut {NoStop}%
\bibitem [{\citenamefont {Callan}\ and\ \citenamefont
  {Coleman}(1977)}]{Callan:1977}%
  \BibitemOpen
  \bibfield  {author} {\bibinfo {author} {\bibfnamefont {Curtis~G.}\
  \bibnamefont {Callan}}\ and\ \bibinfo {author} {\bibfnamefont {Sidney~R.}\
  \bibnamefont {Coleman}},\ }\bibfield  {title} {\enquote {\bibinfo {title}
  {{The Fate of the False Vacuum. 2. First Quantum Corrections}},}\ }\href
  {\doibase 10.1103/PhysRevD.16.1762} {\bibfield  {journal} {\bibinfo
  {journal} {Phys. Rev.}\ }\textbf {\bibinfo {volume} {D16}},\ \bibinfo {pages}
  {1762--1768} (\bibinfo {year} {1977})}\BibitemShut {NoStop}%
\bibitem [{\citenamefont {Coleman}(1988)}]{Coleman:1988}%
  \BibitemOpen
  \bibfield  {author} {\bibinfo {author} {\bibfnamefont {Sidney~R.}\
  \bibnamefont {Coleman}},\ }\bibfield  {title} {\enquote {\bibinfo {title}
  {{Quantum Tunneling and Negative Eigenvalues}},}\ }\href {\doibase
  10.1016/0550-3213(88)90308-2} {\bibfield  {journal} {\bibinfo  {journal}
  {Nucl. Phys.}\ }\textbf {\bibinfo {volume} {B298}},\ \bibinfo {pages}
  {178--186} (\bibinfo {year} {1988})}\BibitemShut {NoStop}%
\bibitem [{\citenamefont {Coleman}\ and\ \citenamefont
  {De~Luccia}(1980)}]{Coleman:1980aw}%
  \BibitemOpen
  \bibfield  {author} {\bibinfo {author} {\bibfnamefont {Sidney~R.}\
  \bibnamefont {Coleman}}\ and\ \bibinfo {author} {\bibfnamefont {Frank}\
  \bibnamefont {De~Luccia}},\ }\bibfield  {title} {\enquote {\bibinfo {title}
  {{Gravitational Effects on and of Vacuum Decay}},}\ }\href {\doibase
  10.1103/PhysRevD.21.3305} {\bibfield  {journal} {\bibinfo  {journal} {Phys.
  Rev.}\ }\textbf {\bibinfo {volume} {D21}},\ \bibinfo {pages} {3305} (\bibinfo
  {year} {1980})}\BibitemShut {NoStop}%
\bibitem [{\citenamefont {Lavrelashvili}\ \emph {et~al.}(1985)\citenamefont
  {Lavrelashvili}, \citenamefont {Rubakov},\ and\ \citenamefont
  {Tinyakov}}]{Lavrelashvili:1985vn}%
  \BibitemOpen
  \bibfield  {author} {\bibinfo {author} {\bibfnamefont {George~V.}\
  \bibnamefont {Lavrelashvili}}, \bibinfo {author} {\bibfnamefont {V.~A.}\
  \bibnamefont {Rubakov}}, \ and\ \bibinfo {author} {\bibfnamefont {P.~G.}\
  \bibnamefont {Tinyakov}},\ }\bibfield  {title} {\enquote {\bibinfo {title}
  {{Tunneling transitions with gravitation: breaking of the quasiclassical
  approximation}},}\ }\href {\doibase 10.1016/0370-2693(85)90761-0} {\bibfield
  {journal} {\bibinfo  {journal} {Phys. Lett.}\ }\textbf {\bibinfo {volume}
  {161B}},\ \bibinfo {pages} {280--284} (\bibinfo {year} {1985})}\BibitemShut
  {NoStop}%
\bibitem [{\citenamefont {Tanaka}\ and\ \citenamefont
  {Sasaki}(1992)}]{Tanaka:1992zw}%
  \BibitemOpen
  \bibfield  {author} {\bibinfo {author} {\bibfnamefont {Takahiro}\
  \bibnamefont {Tanaka}}\ and\ \bibinfo {author} {\bibfnamefont {Misao}\
  \bibnamefont {Sasaki}},\ }\bibfield  {title} {\enquote {\bibinfo {title}
  {{False vacuum decay with gravity: Negative mode problem}},}\ }\href
  {\doibase 10.1143/PTP.88.503} {\bibfield  {journal} {\bibinfo  {journal}
  {Prog. Theor. Phys.}\ }\textbf {\bibinfo {volume} {88}},\ \bibinfo {pages}
  {503--528} (\bibinfo {year} {1992})}\BibitemShut {NoStop}%
\bibitem [{\citenamefont {Garriga}(1994)}]{Garriga:1993fh}%
  \BibitemOpen
  \bibfield  {author} {\bibinfo {author} {\bibfnamefont {Jaume}\ \bibnamefont
  {Garriga}},\ }\bibfield  {title} {\enquote {\bibinfo {title} {{Nucleation
  rates in flat and curved space}},}\ }\href {\doibase
  10.1103/PhysRevD.49.6327} {\bibfield  {journal} {\bibinfo  {journal} {Phys.
  Rev.}\ }\textbf {\bibinfo {volume} {D49}},\ \bibinfo {pages} {6327--6342}
  (\bibinfo {year} {1994})},\ \Eprint {http://arxiv.org/abs/hep-ph/9308280}
  {arXiv:hep-ph/9308280 [hep-ph]} \BibitemShut {NoStop}%
\bibitem [{\citenamefont {Gratton}\ and\ \citenamefont
  {Turok}(1999)}]{Gratton:1999ya}%
  \BibitemOpen
  \bibfield  {author} {\bibinfo {author} {\bibfnamefont {Steven}\ \bibnamefont
  {Gratton}}\ and\ \bibinfo {author} {\bibfnamefont {Neil}\ \bibnamefont
  {Turok}},\ }\bibfield  {title} {\enquote {\bibinfo {title} {{Cosmological
  perturbations from the no boundary Euclidean path integral}},}\ }\href
  {\doibase 10.1103/PhysRevD.60.123507} {\bibfield  {journal} {\bibinfo
  {journal} {Phys. Rev.}\ }\textbf {\bibinfo {volume} {D60}},\ \bibinfo {pages}
  {123507} (\bibinfo {year} {1999})},\ \Eprint
  {http://arxiv.org/abs/astro-ph/9902265} {arXiv:astro-ph/9902265 [astro-ph]}
  \BibitemShut {NoStop}%
\bibitem [{\citenamefont {Lavrelashvili}(2000)}]{Lavrelashvili:1999sr}%
  \BibitemOpen
  \bibfield  {author} {\bibinfo {author} {\bibfnamefont {George~V.}\
  \bibnamefont {Lavrelashvili}},\ }\bibfield  {title} {\enquote {\bibinfo
  {title} {{Negative mode problem in false vacuum decay with gravity}},}\
  }\bibfield  {booktitle} {\emph {\bibinfo {booktitle} {{Constrained dynamics
  and quantum gravity. Proceedings, 3rd Meeting, QG'99, Villasimius, Italy,
  September 13-17, 1999}}},\ }\href {\doibase 10.1016/S0920-5632(00)00756-8}
  {\bibfield  {journal} {\bibinfo  {journal} {Nucl. Phys. Proc. Suppl.}\
  }\textbf {\bibinfo {volume} {88}},\ \bibinfo {pages} {75--82} (\bibinfo
  {year} {2000})},\ \Eprint {http://arxiv.org/abs/gr-qc/0004025}
  {arXiv:gr-qc/0004025 [gr-qc]} \BibitemShut {NoStop}%
\bibitem [{\citenamefont {Tanaka}(1999)}]{Tanaka:1999pj}%
  \BibitemOpen
  \bibfield  {author} {\bibinfo {author} {\bibfnamefont {Takahiro}\
  \bibnamefont {Tanaka}},\ }\bibfield  {title} {\enquote {\bibinfo {title}
  {{The No - negative mode theorem in false vacuum decay with gravity}},}\
  }\href {\doibase 10.1016/S0550-3213(99)00369-7} {\bibfield  {journal}
  {\bibinfo  {journal} {Nucl. Phys.}\ }\textbf {\bibinfo {volume} {B556}},\
  \bibinfo {pages} {373--396} (\bibinfo {year} {1999})},\ \Eprint
  {http://arxiv.org/abs/gr-qc/9901082} {arXiv:gr-qc/9901082 [gr-qc]}
  \BibitemShut {NoStop}%
\bibitem [{\citenamefont {Gratton}\ and\ \citenamefont
  {Turok}(2001)}]{Gratton:2000fj}%
  \BibitemOpen
  \bibfield  {author} {\bibinfo {author} {\bibfnamefont {Steven}\ \bibnamefont
  {Gratton}}\ and\ \bibinfo {author} {\bibfnamefont {Neil}\ \bibnamefont
  {Turok}},\ }\bibfield  {title} {\enquote {\bibinfo {title} {{Homogeneous
  modes of cosmological instantons}},}\ }\href {\doibase
  10.1103/PhysRevD.63.123514} {\bibfield  {journal} {\bibinfo  {journal} {Phys.
  Rev.}\ }\textbf {\bibinfo {volume} {D63}},\ \bibinfo {pages} {123514}
  (\bibinfo {year} {2001})},\ \Eprint {http://arxiv.org/abs/hep-th/0008235}
  {arXiv:hep-th/0008235 [hep-th]} \BibitemShut {NoStop}%
\bibitem [{\citenamefont {Khvedelidze}\ \emph {et~al.}(2000)\citenamefont
  {Khvedelidze}, \citenamefont {Lavrelashvili},\ and\ \citenamefont
  {Tanaka}}]{Khvedelidze:2000cp}%
  \BibitemOpen
  \bibfield  {author} {\bibinfo {author} {\bibfnamefont {Arsen}\ \bibnamefont
  {Khvedelidze}}, \bibinfo {author} {\bibfnamefont {George~V.}\ \bibnamefont
  {Lavrelashvili}}, \ and\ \bibinfo {author} {\bibfnamefont {Takahiro}\
  \bibnamefont {Tanaka}},\ }\bibfield  {title} {\enquote {\bibinfo {title} {{On
  cosmological perturbations in closed FRW model with scalar field and false
  vacuum decay}},}\ }\href {\doibase 10.1103/PhysRevD.62.083501} {\bibfield
  {journal} {\bibinfo  {journal} {Phys. Rev.}\ }\textbf {\bibinfo {volume}
  {D62}},\ \bibinfo {pages} {083501} (\bibinfo {year} {2000})},\ \Eprint
  {http://arxiv.org/abs/gr-qc/0001041} {arXiv:gr-qc/0001041 [gr-qc]}
  \BibitemShut {NoStop}%
\bibitem [{\citenamefont {Dunne}\ and\ \citenamefont
  {Wang}(2006)}]{Dunne:2006bt}%
  \BibitemOpen
  \bibfield  {author} {\bibinfo {author} {\bibfnamefont {Gerald~V.}\
  \bibnamefont {Dunne}}\ and\ \bibinfo {author} {\bibfnamefont {Qing-hai}\
  \bibnamefont {Wang}},\ }\bibfield  {title} {\enquote {\bibinfo {title}
  {{Fluctuations about Cosmological Instantons}},}\ }\href {\doibase
  10.1103/PhysRevD.74.024018} {\bibfield  {journal} {\bibinfo  {journal} {Phys.
  Rev.}\ }\textbf {\bibinfo {volume} {D74}},\ \bibinfo {pages} {024018}
  (\bibinfo {year} {2006})},\ \Eprint {http://arxiv.org/abs/hep-th/0605176}
  {arXiv:hep-th/0605176 [hep-th]} \BibitemShut {NoStop}%
\bibitem [{\citenamefont {Lee}\ and\ \citenamefont
  {Weinberg}(2014)}]{Lee:2014uza}%
  \BibitemOpen
  \bibfield  {author} {\bibinfo {author} {\bibfnamefont {Hakjoon}\ \bibnamefont
  {Lee}}\ and\ \bibinfo {author} {\bibfnamefont {Erick~J.}\ \bibnamefont
  {Weinberg}},\ }\bibfield  {title} {\enquote {\bibinfo {title} {{Negative
  modes of Coleman-De Luccia bounces}},}\ }\href {\doibase
  10.1103/PhysRevD.90.124002} {\bibfield  {journal} {\bibinfo  {journal} {Phys.
  Rev.}\ }\textbf {\bibinfo {volume} {D90}},\ \bibinfo {pages} {124002}
  (\bibinfo {year} {2014})},\ \Eprint {http://arxiv.org/abs/1408.6547}
  {arXiv:1408.6547 [hep-th]} \BibitemShut {NoStop}%
\bibitem [{\citenamefont {Koehn}\ \emph {et~al.}(2015)\citenamefont {Koehn},
  \citenamefont {Lavrelashvili},\ and\ \citenamefont
  {Lehners}}]{Koehn:2015hga}%
  \BibitemOpen
  \bibfield  {author} {\bibinfo {author} {\bibfnamefont {Michael}\ \bibnamefont
  {Koehn}}, \bibinfo {author} {\bibfnamefont {George}\ \bibnamefont
  {Lavrelashvili}}, \ and\ \bibinfo {author} {\bibfnamefont {Jean-Luc}\
  \bibnamefont {Lehners}},\ }\bibfield  {title} {\enquote {\bibinfo {title}
  {{Towards a Solution of the Negative Mode Problem in Quantum Tunnelling with
  Gravity}},}\ }\href {\doibase 10.1103/PhysRevD.92.023506} {\bibfield
  {journal} {\bibinfo  {journal} {Phys. Rev.}\ }\textbf {\bibinfo {volume}
  {D92}},\ \bibinfo {pages} {023506} (\bibinfo {year} {2015})},\ \Eprint
  {http://arxiv.org/abs/1504.04334} {arXiv:1504.04334 [hep-th]} \BibitemShut
  {NoStop}%
\bibitem [{\citenamefont {Krive}\ and\ \citenamefont
  {Linde}(1976)}]{Krive:1977}%
  \BibitemOpen
  \bibfield  {author} {\bibinfo {author} {\bibfnamefont {Ilya~V.}\ \bibnamefont
  {Krive}}\ and\ \bibinfo {author} {\bibfnamefont {Andrei~D.}\ \bibnamefont
  {Linde}},\ }\bibfield  {title} {\enquote {\bibinfo {title} {{On the Vacuum
  stability problem in gauge theories}},}\ }\href {\doibase
  10.1016/0370-2693(78)90525-7} {\bibfield  {journal} {\bibinfo  {journal}
  {Nucl. Phys.}\ }\textbf {\bibinfo {volume} {B432}},\ \bibinfo {pages} {265}
  (\bibinfo {year} {1976})}\BibitemShut {NoStop}%
\bibitem [{\citenamefont {Cabibbo}\ \emph {et~al.}(1979)\citenamefont
  {Cabibbo}, \citenamefont {Maiani}, \citenamefont {Parisi},\ and\
  \citenamefont {Petronzio}}]{Cabibbo:1979ay}%
  \BibitemOpen
  \bibfield  {author} {\bibinfo {author} {\bibfnamefont {Nicola}\ \bibnamefont
  {Cabibbo}}, \bibinfo {author} {\bibfnamefont {Luciano}\ \bibnamefont
  {Maiani}}, \bibinfo {author} {\bibfnamefont {Giorgio}\ \bibnamefont
  {Parisi}}, \ and\ \bibinfo {author} {\bibfnamefont {Roberto}\ \bibnamefont
  {Petronzio}},\ }\bibfield  {title} {\enquote {\bibinfo {title} {{Bounds on
  the Fermions and Higgs Boson Masses in Grand Unified Theories}},}\ }\href
  {\doibase 10.1016/0550-3213(79)90167-6} {\bibfield  {journal} {\bibinfo
  {journal} {Nucl. Phys.}\ }\textbf {\bibinfo {volume} {B158}},\ \bibinfo
  {pages} {295--305} (\bibinfo {year} {1979})}\BibitemShut {NoStop}%
\bibitem [{\citenamefont {Politzer}\ and\ \citenamefont
  {Wolfram}(1979)}]{Polizer:1979}%
  \BibitemOpen
  \bibfield  {author} {\bibinfo {author} {\bibfnamefont {H.~David}\
  \bibnamefont {Politzer}}\ and\ \bibinfo {author} {\bibfnamefont {Stephen}\
  \bibnamefont {Wolfram}},\ }\bibfield  {title} {\enquote {\bibinfo {title}
  {{Bounds on particle masses in the Weinberg-Salam model}},}\ }\href {\doibase
  http://dx.doi.org/10.1016/0370-2693(79)90746-9} {\bibfield  {journal}
  {\bibinfo  {journal} {Physics Letters B}\ }\textbf {\bibinfo {volume} {82}},\
  \bibinfo {pages} {242 -- 246} (\bibinfo {year} {1979})}\BibitemShut {NoStop}%
\bibitem [{\citenamefont {Isidori}\ \emph {et~al.}(2001)\citenamefont
  {Isidori}, \citenamefont {Ridolfi},\ and\ \citenamefont
  {Strumia}}]{Isidori:2001bm}%
  \BibitemOpen
  \bibfield  {author} {\bibinfo {author} {\bibfnamefont {Gino}\ \bibnamefont
  {Isidori}}, \bibinfo {author} {\bibfnamefont {Giovanni}\ \bibnamefont
  {Ridolfi}}, \ and\ \bibinfo {author} {\bibfnamefont {Alessandro}\
  \bibnamefont {Strumia}},\ }\bibfield  {title} {\enquote {\bibinfo {title}
  {{On the metastability of the standard model vacuum}},}\ }\href {\doibase
  10.1016/S0550-3213(01)00302-9} {\bibfield  {journal} {\bibinfo  {journal}
  {Nucl. Phys.}\ }\textbf {\bibinfo {volume} {B609}},\ \bibinfo {pages}
  {387--409} (\bibinfo {year} {2001})},\ \Eprint
  {http://arxiv.org/abs/hep-ph/0104016} {arXiv:hep-ph/0104016 [hep-ph]}
  \BibitemShut {NoStop}%
\bibitem [{\citenamefont {Rajantie}\ and\ \citenamefont
  {Stopyra}(2017{\natexlab{a}})}]{Rajantie:2016hkj}%
  \BibitemOpen
  \bibfield  {author} {\bibinfo {author} {\bibfnamefont {Arttu}\ \bibnamefont
  {Rajantie}}\ and\ \bibinfo {author} {\bibfnamefont {Stephen}\ \bibnamefont
  {Stopyra}},\ }\bibfield  {title} {\enquote {\bibinfo {title} {{Standard Model
  vacuum decay with gravity}},}\ }\href {\doibase 10.1103/PhysRevD.95.025008}
  {\bibfield  {journal} {\bibinfo  {journal} {Phys. Rev.}\ }\textbf {\bibinfo
  {volume} {D95}},\ \bibinfo {pages} {025008} (\bibinfo {year}
  {2017}{\natexlab{a}})},\ \Eprint {http://arxiv.org/abs/1606.00849}
  {arXiv:1606.00849 [hep-th]} \BibitemShut {NoStop}%
\bibitem [{\citenamefont {Andreassen}\ \emph {et~al.}(2018)\citenamefont
  {Andreassen}, \citenamefont {Frost},\ and\ \citenamefont
  {Schwartz}}]{Andreassen:2017rzq}%
  \BibitemOpen
  \bibfield  {author} {\bibinfo {author} {\bibfnamefont {Anders}\ \bibnamefont
  {Andreassen}}, \bibinfo {author} {\bibfnamefont {William}\ \bibnamefont
  {Frost}}, \ and\ \bibinfo {author} {\bibfnamefont {Matthew~D.}\ \bibnamefont
  {Schwartz}},\ }\bibfield  {title} {\enquote {\bibinfo {title} {{Scale
  Invariant Instantons and the Complete Lifetime of the Standard Model}},}\
  }\href {\doibase 10.1103/PhysRevD.97.056006} {\bibfield  {journal} {\bibinfo
  {journal} {Phys. Rev.}\ }\textbf {\bibinfo {volume} {D97}},\ \bibinfo {pages}
  {056006} (\bibinfo {year} {2018})},\ \Eprint
  {http://arxiv.org/abs/1707.08124} {arXiv:1707.08124 [hep-ph]} \BibitemShut
  {NoStop}%
\bibitem [{\citenamefont {Branchina}\ \emph {et~al.}(2018)\citenamefont
  {Branchina}, \citenamefont {Contino},\ and\ \citenamefont
  {Pilaftsis}}]{Branchina:2018xdh}%
  \BibitemOpen
  \bibfield  {author} {\bibinfo {author} {\bibfnamefont {Vincenzo}\
  \bibnamefont {Branchina}}, \bibinfo {author} {\bibfnamefont {Filippo}\
  \bibnamefont {Contino}}, \ and\ \bibinfo {author} {\bibfnamefont {Apostolos}\
  \bibnamefont {Pilaftsis}},\ }\bibfield  {title} {\enquote {\bibinfo {title}
  {{Protecting the Stability of the EW Vacuum from Planck-Scale Gravitational
  Effects}},}\ }\href@noop {} {\  (\bibinfo {year} {2018})},\ \Eprint
  {http://arxiv.org/abs/1806.11059} {arXiv:1806.11059 [hep-ph]} \BibitemShut
  {NoStop}%
\bibitem [{\citenamefont {Degrassi}\ \emph {et~al.}(2012)\citenamefont
  {Degrassi}, \citenamefont {Di~Vita}, \citenamefont {Elias-Miro},
  \citenamefont {Espinosa}, \citenamefont {Giudice}, \citenamefont {Isidori},\
  and\ \citenamefont {Strumia}}]{Degrassi:2012ry}%
  \BibitemOpen
  \bibfield  {author} {\bibinfo {author} {\bibfnamefont {Giuseppe}\
  \bibnamefont {Degrassi}}, \bibinfo {author} {\bibfnamefont {Stefano}\
  \bibnamefont {Di~Vita}}, \bibinfo {author} {\bibfnamefont {Joan}\
  \bibnamefont {Elias-Miro}}, \bibinfo {author} {\bibfnamefont {Jose~R.}\
  \bibnamefont {Espinosa}}, \bibinfo {author} {\bibfnamefont {Gian~F.}\
  \bibnamefont {Giudice}}, \bibinfo {author} {\bibfnamefont {Gino}\
  \bibnamefont {Isidori}}, \ and\ \bibinfo {author} {\bibfnamefont
  {Alessandro}\ \bibnamefont {Strumia}},\ }\bibfield  {title} {\enquote
  {\bibinfo {title} {{Higgs mass and vacuum stability in the Standard Model at
  NNLO}},}\ }\href {\doibase 10.1007/JHEP08(2012)098} {\bibfield  {journal}
  {\bibinfo  {journal} {JHEP}\ }\textbf {\bibinfo {volume} {08}},\ \bibinfo
  {pages} {098} (\bibinfo {year} {2012})},\ \Eprint
  {http://arxiv.org/abs/1205.6497} {arXiv:1205.6497 [hep-ph]} \BibitemShut
  {NoStop}%
\bibitem [{\citenamefont {Buttazzo}\ \emph {et~al.}(2013)\citenamefont
  {Buttazzo}, \citenamefont {Degrassi}, \citenamefont {Giardino}, \citenamefont
  {Giudice}, \citenamefont {Sala}, \citenamefont {Salvio},\ and\ \citenamefont
  {Strumia}}]{Buttazzo:2013uya}%
  \BibitemOpen
  \bibfield  {author} {\bibinfo {author} {\bibfnamefont {Dario}\ \bibnamefont
  {Buttazzo}}, \bibinfo {author} {\bibfnamefont {Giuseppe}\ \bibnamefont
  {Degrassi}}, \bibinfo {author} {\bibfnamefont {Pier~Paolo}\ \bibnamefont
  {Giardino}}, \bibinfo {author} {\bibfnamefont {Gian~F.}\ \bibnamefont
  {Giudice}}, \bibinfo {author} {\bibfnamefont {Filippo}\ \bibnamefont {Sala}},
  \bibinfo {author} {\bibfnamefont {Alberto}\ \bibnamefont {Salvio}}, \ and\
  \bibinfo {author} {\bibfnamefont {Alessandro}\ \bibnamefont {Strumia}},\
  }\bibfield  {title} {\enquote {\bibinfo {title} {{Investigating the
  near-criticality of the Higgs boson}},}\ }\href {\doibase
  10.1007/JHEP12(2013)089} {\bibfield  {journal} {\bibinfo  {journal} {JHEP}\
  }\textbf {\bibinfo {volume} {12}},\ \bibinfo {pages} {089} (\bibinfo {year}
  {2013})},\ \Eprint {http://arxiv.org/abs/1307.3536} {arXiv:1307.3536
  [hep-ph]} \BibitemShut {NoStop}%
\bibitem [{\citenamefont {Blum}\ \emph {et~al.}(2015)\citenamefont {Blum},
  \citenamefont {D'Agnolo},\ and\ \citenamefont {Fan}}]{Blum:2015rpa}%
  \BibitemOpen
  \bibfield  {author} {\bibinfo {author} {\bibfnamefont {Kfir}\ \bibnamefont
  {Blum}}, \bibinfo {author} {\bibfnamefont {Raffaele~Tito}\ \bibnamefont
  {D'Agnolo}}, \ and\ \bibinfo {author} {\bibfnamefont {JiJi}\ \bibnamefont
  {Fan}},\ }\bibfield  {title} {\enquote {\bibinfo {title} {{Vacuum stability
  bounds on Higgs coupling deviations in the absence of new bosons}},}\ }\href
  {\doibase 10.1007/JHEP03(2015)166} {\bibfield  {journal} {\bibinfo  {journal}
  {JHEP}\ }\textbf {\bibinfo {volume} {03}},\ \bibinfo {pages} {166} (\bibinfo
  {year} {2015})},\ \Eprint {http://arxiv.org/abs/1502.01045} {arXiv:1502.01045
  [hep-ph]} \BibitemShut {NoStop}%
\bibitem [{\citenamefont {Gregory}\ \emph {et~al.}(2014)\citenamefont
  {Gregory}, \citenamefont {Moss},\ and\ \citenamefont
  {Withers}}]{Gregory:2013hja}%
  \BibitemOpen
  \bibfield  {author} {\bibinfo {author} {\bibfnamefont {Ruth}\ \bibnamefont
  {Gregory}}, \bibinfo {author} {\bibfnamefont {Ian}\ \bibnamefont {Moss}}, \
  and\ \bibinfo {author} {\bibfnamefont {Benjamin}\ \bibnamefont {Withers}},\
  }\bibfield  {title} {\enquote {\bibinfo {title} {{Black holes as bubble
  nucleation sites}},}\ }\href {\doibase 10.1007/JHEP03(2014)081} {\bibfield
  {journal} {\bibinfo  {journal} {JHEP}\ }\textbf {\bibinfo {volume} {03}},\
  \bibinfo {pages} {081} (\bibinfo {year} {2014})},\ \Eprint
  {http://arxiv.org/abs/1401.0017} {arXiv:1401.0017 [hep-th]} \BibitemShut
  {NoStop}%
\bibitem [{\citenamefont {Burda}\ \emph
  {et~al.}(2015{\natexlab{a}})\citenamefont {Burda}, \citenamefont {Gregory},\
  and\ \citenamefont {Moss}}]{Burda:2015isa}%
  \BibitemOpen
  \bibfield  {author} {\bibinfo {author} {\bibfnamefont {Philipp}\ \bibnamefont
  {Burda}}, \bibinfo {author} {\bibfnamefont {Ruth}\ \bibnamefont {Gregory}}, \
  and\ \bibinfo {author} {\bibfnamefont {Ian}\ \bibnamefont {Moss}},\
  }\bibfield  {title} {\enquote {\bibinfo {title} {{Gravity and the stability
  of the Higgs vacuum}},}\ }\href {\doibase 10.1103/PhysRevLett.115.071303}
  {\bibfield  {journal} {\bibinfo  {journal} {Phys. Rev. Lett.}\ }\textbf
  {\bibinfo {volume} {115}},\ \bibinfo {pages} {071303} (\bibinfo {year}
  {2015}{\natexlab{a}})},\ \Eprint {http://arxiv.org/abs/1501.04937}
  {arXiv:1501.04937 [hep-th]} \BibitemShut {NoStop}%
\bibitem [{\citenamefont {Burda}\ \emph
  {et~al.}(2015{\natexlab{b}})\citenamefont {Burda}, \citenamefont {Gregory},\
  and\ \citenamefont {Moss}}]{Burda:2015yfa}%
  \BibitemOpen
  \bibfield  {author} {\bibinfo {author} {\bibfnamefont {Philipp}\ \bibnamefont
  {Burda}}, \bibinfo {author} {\bibfnamefont {Ruth}\ \bibnamefont {Gregory}}, \
  and\ \bibinfo {author} {\bibfnamefont {Ian}\ \bibnamefont {Moss}},\
  }\bibfield  {title} {\enquote {\bibinfo {title} {{Vacuum metastability with
  black holes}},}\ }\href {\doibase 10.1007/JHEP08(2015)114} {\bibfield
  {journal} {\bibinfo  {journal} {JHEP}\ }\textbf {\bibinfo {volume} {08}},\
  \bibinfo {pages} {114} (\bibinfo {year} {2015}{\natexlab{b}})},\ \Eprint
  {http://arxiv.org/abs/1503.07331} {arXiv:1503.07331 [hep-th]} \BibitemShut
  {NoStop}%
\bibitem [{\citenamefont {Burda}\ \emph {et~al.}(2016)\citenamefont {Burda},
  \citenamefont {Gregory},\ and\ \citenamefont {Moss}}]{Burda:2016mou}%
  \BibitemOpen
  \bibfield  {author} {\bibinfo {author} {\bibfnamefont {Philipp}\ \bibnamefont
  {Burda}}, \bibinfo {author} {\bibfnamefont {Ruth}\ \bibnamefont {Gregory}}, \
  and\ \bibinfo {author} {\bibfnamefont {Ian}\ \bibnamefont {Moss}},\
  }\bibfield  {title} {\enquote {\bibinfo {title} {{The fate of the Higgs
  vacuum}},}\ }\href {\doibase 10.1007/JHEP06(2016)025} {\bibfield  {journal}
  {\bibinfo  {journal} {JHEP}\ }\textbf {\bibinfo {volume} {06}},\ \bibinfo
  {pages} {025} (\bibinfo {year} {2016})},\ \Eprint
  {http://arxiv.org/abs/1601.02152} {arXiv:1601.02152 [hep-th]} \BibitemShut
  {NoStop}%
\bibitem [{\citenamefont {Gregory}\ and\ \citenamefont
  {Moss}(2016)}]{Gregory:2016xix}%
  \BibitemOpen
  \bibfield  {author} {\bibinfo {author} {\bibfnamefont {Ruth}\ \bibnamefont
  {Gregory}}\ and\ \bibinfo {author} {\bibfnamefont {Ian~G.}\ \bibnamefont
  {Moss}},\ }\bibfield  {title} {\enquote {\bibinfo {title} {{The Fate of the
  Higgs Vacuum}},}\ }\bibfield  {booktitle} {\emph {\bibinfo {booktitle}
  {{Proceedings, 38th International Conference on High Energy Physics (ICHEP
  2016): Chicago, IL, USA, August 3-10, 2016}}},\ }\href@noop {} {\bibfield
  {journal} {\bibinfo  {journal} {PoS}\ }\textbf {\bibinfo {volume}
  {ICHEP2016}},\ \bibinfo {pages} {344} (\bibinfo {year} {2016})},\ \Eprint
  {http://arxiv.org/abs/1611.04935} {arXiv:1611.04935 [hep-th]} \BibitemShut
  {NoStop}%
\bibitem [{\citenamefont {Tetradis}(2016)}]{Tetradis:2016vqb}%
  \BibitemOpen
  \bibfield  {author} {\bibinfo {author} {\bibfnamefont {Nikolaos}\
  \bibnamefont {Tetradis}},\ }\bibfield  {title} {\enquote {\bibinfo {title}
  {{Black holes and Higgs stability}},}\ }\href {\doibase
  10.1088/1475-7516/2016/09/036} {\bibfield  {journal} {\bibinfo  {journal}
  {JCAP}\ }\textbf {\bibinfo {volume} {1609}},\ \bibinfo {pages} {036}
  (\bibinfo {year} {2016})},\ \Eprint {http://arxiv.org/abs/1606.04018}
  {arXiv:1606.04018 [hep-ph]} \BibitemShut {NoStop}%
\bibitem [{\citenamefont {Chen}\ \emph {et~al.}(2017)\citenamefont {Chen},
  \citenamefont {Domènech}, \citenamefont {Sasaki},\ and\ \citenamefont
  {Yeom}}]{Chen:2017suz}%
  \BibitemOpen
  \bibfield  {author} {\bibinfo {author} {\bibfnamefont {Pisin}\ \bibnamefont
  {Chen}}, \bibinfo {author} {\bibfnamefont {Guillem}\ \bibnamefont
  {Domènech}}, \bibinfo {author} {\bibfnamefont {Misao}\ \bibnamefont
  {Sasaki}}, \ and\ \bibinfo {author} {\bibfnamefont {Dong-han}\ \bibnamefont
  {Yeom}},\ }\bibfield  {title} {\enquote {\bibinfo {title} {{Thermal
  activation of thin-shells in anti-de Sitter black hole spacetime}},}\ }\href
  {\doibase 10.1007/JHEP07(2017)134} {\bibfield  {journal} {\bibinfo  {journal}
  {JHEP}\ }\textbf {\bibinfo {volume} {07}},\ \bibinfo {pages} {134} (\bibinfo
  {year} {2017})},\ \Eprint {http://arxiv.org/abs/1704.04020} {arXiv:1704.04020
  [gr-qc]} \BibitemShut {NoStop}%
\bibitem [{\citenamefont {Mukaida}\ and\ \citenamefont
  {Yamada}(2017)}]{Mukaida:2017bgd}%
  \BibitemOpen
  \bibfield  {author} {\bibinfo {author} {\bibfnamefont {Kyohei}\ \bibnamefont
  {Mukaida}}\ and\ \bibinfo {author} {\bibfnamefont {Masaki}\ \bibnamefont
  {Yamada}},\ }\bibfield  {title} {\enquote {\bibinfo {title} {{False Vacuum
  Decay Catalyzed by Black Holes}},}\ }\href {\doibase
  10.1103/PhysRevD.96.103514} {\bibfield  {journal} {\bibinfo  {journal} {Phys.
  Rev.}\ }\textbf {\bibinfo {volume} {D96}},\ \bibinfo {pages} {103514}
  (\bibinfo {year} {2017})},\ \Eprint {http://arxiv.org/abs/1706.04523}
  {arXiv:1706.04523 [hep-th]} \BibitemShut {NoStop}%
\bibitem [{\citenamefont {Gorbunov}\ \emph {et~al.}(2017)\citenamefont
  {Gorbunov}, \citenamefont {Levkov},\ and\ \citenamefont
  {Panin}}]{Gorbunov:2017fhq}%
  \BibitemOpen
  \bibfield  {author} {\bibinfo {author} {\bibfnamefont {Dmitry}\ \bibnamefont
  {Gorbunov}}, \bibinfo {author} {\bibfnamefont {Dmitry}\ \bibnamefont
  {Levkov}}, \ and\ \bibinfo {author} {\bibfnamefont {Alexander}\ \bibnamefont
  {Panin}},\ }\bibfield  {title} {\enquote {\bibinfo {title} {{Fatal youth of
  the Universe: black hole threat for the electroweak vacuum during
  preheating}},}\ }\href {\doibase 10.1088/1475-7516/2017/10/016} {\bibfield
  {journal} {\bibinfo  {journal} {JCAP}\ }\textbf {\bibinfo {volume} {1710}},\
  \bibinfo {pages} {016} (\bibinfo {year} {2017})},\ \Eprint
  {http://arxiv.org/abs/1704.05399} {arXiv:1704.05399 [astro-ph.CO]}
  \BibitemShut {NoStop}%
\bibitem [{\citenamefont {Chen}\ \emph {et~al.}(2018)\citenamefont {Chen},
  \citenamefont {Sasaki},\ and\ \citenamefont {Yeom}}]{Chen:2018aij}%
  \BibitemOpen
  \bibfield  {author} {\bibinfo {author} {\bibfnamefont {Pisin}\ \bibnamefont
  {Chen}}, \bibinfo {author} {\bibfnamefont {Misao}\ \bibnamefont {Sasaki}}, \
  and\ \bibinfo {author} {\bibfnamefont {Dong-Han}\ \bibnamefont {Yeom}},\
  }\bibfield  {title} {\enquote {\bibinfo {title} {{Hawking radiation as
  instantons}},}\ }\href@noop {} {\  (\bibinfo {year} {2018})},\ \Eprint
  {http://arxiv.org/abs/1806.03766} {arXiv:1806.03766 [hep-th]} \BibitemShut
  {NoStop}%
\bibitem [{\citenamefont {Aguirre}\ and\ \citenamefont
  {Johnson}(2006)}]{Aguirre:2005nt}%
  \BibitemOpen
  \bibfield  {author} {\bibinfo {author} {\bibfnamefont {Anthony}\ \bibnamefont
  {Aguirre}}\ and\ \bibinfo {author} {\bibfnamefont {Matthew~C.}\ \bibnamefont
  {Johnson}},\ }\bibfield  {title} {\enquote {\bibinfo {title} {{Two tunnels to
  inflation}},}\ }\href {\doibase 10.1103/PhysRevD.73.123529} {\bibfield
  {journal} {\bibinfo  {journal} {Phys. Rev.}\ }\textbf {\bibinfo {volume}
  {D73}},\ \bibinfo {pages} {123529} (\bibinfo {year} {2006})},\ \Eprint
  {http://arxiv.org/abs/gr-qc/0512034} {arXiv:gr-qc/0512034 [gr-qc]}
  \BibitemShut {NoStop}%
\bibitem [{\citenamefont {Aguirre}\ \emph {et~al.}(2006)\citenamefont
  {Aguirre}, \citenamefont {Banks},\ and\ \citenamefont
  {Johnson}}]{Aguirre:2006ap}%
  \BibitemOpen
  \bibfield  {author} {\bibinfo {author} {\bibfnamefont {A.}~\bibnamefont
  {Aguirre}}, \bibinfo {author} {\bibfnamefont {T.}~\bibnamefont {Banks}}, \
  and\ \bibinfo {author} {\bibfnamefont {M.}~\bibnamefont {Johnson}},\
  }\bibfield  {title} {\enquote {\bibinfo {title} {{Regulating eternal
  inflation. II. The Great divide}},}\ }\href {\doibase
  10.1088/1126-6708/2006/08/065} {\bibfield  {journal} {\bibinfo  {journal}
  {JHEP}\ }\textbf {\bibinfo {volume} {08}},\ \bibinfo {pages} {065} (\bibinfo
  {year} {2006})},\ \Eprint {http://arxiv.org/abs/hep-th/0603107}
  {arXiv:hep-th/0603107 [hep-th]} \BibitemShut {NoStop}%
\bibitem [{\citenamefont {Bousso}\ \emph {et~al.}(2006)\citenamefont {Bousso},
  \citenamefont {Freivogel},\ and\ \citenamefont {Lippert}}]{Bousso:2006am}%
  \BibitemOpen
  \bibfield  {author} {\bibinfo {author} {\bibfnamefont {Raphael}\ \bibnamefont
  {Bousso}}, \bibinfo {author} {\bibfnamefont {Ben}\ \bibnamefont {Freivogel}},
  \ and\ \bibinfo {author} {\bibfnamefont {Matthew}\ \bibnamefont {Lippert}},\
  }\bibfield  {title} {\enquote {\bibinfo {title} {{Probabilities in the
  landscape: The Decay of nearly flat space}},}\ }\href {\doibase
  10.1103/PhysRevD.74.046008} {\bibfield  {journal} {\bibinfo  {journal} {Phys.
  Rev.}\ }\textbf {\bibinfo {volume} {D74}},\ \bibinfo {pages} {046008}
  (\bibinfo {year} {2006})},\ \Eprint {http://arxiv.org/abs/hep-th/0603105}
  {arXiv:hep-th/0603105 [hep-th]} \BibitemShut {NoStop}%
\bibitem [{\citenamefont {Isidori}\ \emph {et~al.}(2008)\citenamefont
  {Isidori}, \citenamefont {Rychkov}, \citenamefont {Strumia},\ and\
  \citenamefont {Tetradis}}]{Isidori:2007vm}%
  \BibitemOpen
  \bibfield  {author} {\bibinfo {author} {\bibfnamefont {Gino}\ \bibnamefont
  {Isidori}}, \bibinfo {author} {\bibfnamefont {Vyacheslav~S.}\ \bibnamefont
  {Rychkov}}, \bibinfo {author} {\bibfnamefont {Alessandro}\ \bibnamefont
  {Strumia}}, \ and\ \bibinfo {author} {\bibfnamefont {Nikolaos}\ \bibnamefont
  {Tetradis}},\ }\bibfield  {title} {\enquote {\bibinfo {title} {{Gravitational
  corrections to standard model vacuum decay}},}\ }\href {\doibase
  10.1103/PhysRevD.77.025034} {\bibfield  {journal} {\bibinfo  {journal} {Phys.
  Rev.}\ }\textbf {\bibinfo {volume} {D77}},\ \bibinfo {pages} {025034}
  (\bibinfo {year} {2008})},\ \Eprint {http://arxiv.org/abs/0712.0242}
  {arXiv:0712.0242 [hep-ph]} \BibitemShut {NoStop}%
\bibitem [{\citenamefont {Salvio}\ \emph {et~al.}(2016)\citenamefont {Salvio},
  \citenamefont {Strumia}, \citenamefont {Tetradis},\ and\ \citenamefont
  {Urbano}}]{Salvio:2016mvj}%
  \BibitemOpen
  \bibfield  {author} {\bibinfo {author} {\bibfnamefont {Alberto}\ \bibnamefont
  {Salvio}}, \bibinfo {author} {\bibfnamefont {Alessandro}\ \bibnamefont
  {Strumia}}, \bibinfo {author} {\bibfnamefont {Nikolaos}\ \bibnamefont
  {Tetradis}}, \ and\ \bibinfo {author} {\bibfnamefont {Alfredo}\ \bibnamefont
  {Urbano}},\ }\bibfield  {title} {\enquote {\bibinfo {title} {{On
  gravitational and thermal corrections to vacuum decay}},}\ }\href {\doibase
  10.1007/JHEP09(2016)054} {\bibfield  {journal} {\bibinfo  {journal} {JHEP}\
  }\textbf {\bibinfo {volume} {09}},\ \bibinfo {pages} {054} (\bibinfo {year}
  {2016})},\ \Eprint {http://arxiv.org/abs/1608.02555} {arXiv:1608.02555
  [hep-ph]} \BibitemShut {NoStop}%
\bibitem [{\citenamefont {Rajantie}\ and\ \citenamefont
  {Stopyra}(2017{\natexlab{b}})}]{Rajantie:2017ajw}%
  \BibitemOpen
  \bibfield  {author} {\bibinfo {author} {\bibfnamefont {Arttu}\ \bibnamefont
  {Rajantie}}\ and\ \bibinfo {author} {\bibfnamefont {Stephen}\ \bibnamefont
  {Stopyra}},\ }\bibfield  {title} {\enquote {\bibinfo {title} {{Standard Model
  vacuum decay in a de Sitter Background}},}\ }\href@noop {} {\  (\bibinfo
  {year} {2017}{\natexlab{b}})},\ \Eprint {http://arxiv.org/abs/1707.09175}
  {arXiv:1707.09175 [hep-th]} \BibitemShut {NoStop}%
\bibitem [{\citenamefont {Coleman}\ \emph {et~al.}(1978)\citenamefont
  {Coleman}, \citenamefont {Glaser},\ and\ \citenamefont
  {Martin}}]{Coleman:1977th}%
  \BibitemOpen
  \bibfield  {author} {\bibinfo {author} {\bibfnamefont {Sidney~R.}\
  \bibnamefont {Coleman}}, \bibinfo {author} {\bibfnamefont {V.}~\bibnamefont
  {Glaser}}, \ and\ \bibinfo {author} {\bibfnamefont {Andre}\ \bibnamefont
  {Martin}},\ }\bibfield  {title} {\enquote {\bibinfo {title} {{Action Minima
  Among Solutions to a Class of Euclidean Scalar Field Equations}},}\ }\href
  {\doibase 10.1007/BF01609421} {\bibfield  {journal} {\bibinfo  {journal}
  {Commun. Math. Phys.}\ }\textbf {\bibinfo {volume} {58}},\ \bibinfo {pages}
  {211} (\bibinfo {year} {1978})}\BibitemShut {NoStop}%
\bibitem [{\citenamefont {Garfinkle}\ and\ \citenamefont
  {Gregory}(1990)}]{Garfinkle:1989mv}%
  \BibitemOpen
  \bibfield  {author} {\bibinfo {author} {\bibfnamefont {David}\ \bibnamefont
  {Garfinkle}}\ and\ \bibinfo {author} {\bibfnamefont {Ruth}\ \bibnamefont
  {Gregory}},\ }\bibfield  {title} {\enquote {\bibinfo {title} {{Corrections to
  the Thin Wall Approximation in General Relativity}},}\ }\href {\doibase
  10.1103/PhysRevD.41.1889} {\bibfield  {journal} {\bibinfo  {journal} {Phys.
  Rev.}\ }\textbf {\bibinfo {volume} {D41}},\ \bibinfo {pages} {1889} (\bibinfo
  {year} {1990})}\BibitemShut {NoStop}%
\bibitem [{\citenamefont {{Amann}}\ and\ \citenamefont
  {{Quittner}}(1995)}]{1995JMP....36.4553A}%
  \BibitemOpen
  \bibfield  {author} {\bibinfo {author} {\bibfnamefont {Herbert}\ \bibnamefont
  {{Amann}}}\ and\ \bibinfo {author} {\bibfnamefont {Pavol}\ \bibnamefont
  {{Quittner}}},\ }\bibfield  {title} {\enquote {\bibinfo {title} {{A nodal
  theorem for coupled systems of Schr{\"o}dinger equations and the number of
  bound states}},}\ }\href {\doibase 10.1063/1.530907} {\bibfield  {journal}
  {\bibinfo  {journal} {Journal of Mathematical Physics}\ }\textbf {\bibinfo
  {volume} {36}},\ \bibinfo {pages} {4553--4560} (\bibinfo {year}
  {1995})}\BibitemShut {NoStop}%
\bibitem [{\citenamefont {Hartman}(2002)}]{doi:10.1137/1.9780898719222}%
  \BibitemOpen
  \bibfield  {author} {\bibinfo {author} {\bibfnamefont {Philip}\ \bibnamefont
  {Hartman}},\ }\href {\doibase 10.1137/1.9780898719222} {\emph {\bibinfo
  {title} {Ordinary Differential Equations}}},\ \bibinfo {edition} {2nd}\ ed.\
  (\bibinfo  {publisher} {Society for Industrial and Applied Mathematics},\
  \bibinfo {year} {2002})\BibitemShut {NoStop}%
\bibitem [{\citenamefont {Battarra}\ \emph {et~al.}(2012)\citenamefont
  {Battarra}, \citenamefont {Lavrelashvili},\ and\ \citenamefont
  {Lehners}}]{Battarra:2012vu}%
  \BibitemOpen
  \bibfield  {author} {\bibinfo {author} {\bibfnamefont {Lorenzo}\ \bibnamefont
  {Battarra}}, \bibinfo {author} {\bibfnamefont {George}\ \bibnamefont
  {Lavrelashvili}}, \ and\ \bibinfo {author} {\bibfnamefont {Jean-Luc}\
  \bibnamefont {Lehners}},\ }\bibfield  {title} {\enquote {\bibinfo {title}
  {{Negative Modes of Oscillating Instantons}},}\ }\href {\doibase
  10.1103/PhysRevD.86.124001} {\bibfield  {journal} {\bibinfo  {journal} {Phys.
  Rev.}\ }\textbf {\bibinfo {volume} {D86}},\ \bibinfo {pages} {124001}
  (\bibinfo {year} {2012})},\ \Eprint {http://arxiv.org/abs/1208.2182}
  {arXiv:1208.2182 [hep-th]} \BibitemShut {NoStop}%
\bibitem [{\citenamefont {Gibbons}(1993)}]{Gibbons:1993in}%
  \BibitemOpen
  \bibfield  {author} {\bibinfo {author} {\bibfnamefont {G.~W.}\ \bibnamefont
  {Gibbons}},\ }\bibfield  {title} {\enquote {\bibinfo {title} {{Global
  structure of supergravity domain wall space-times}},}\ }\href {\doibase
  10.1016/0550-3213(93)90099-B} {\bibfield  {journal} {\bibinfo  {journal}
  {Nucl. Phys.}\ }\textbf {\bibinfo {volume} {B394}},\ \bibinfo {pages} {3--20}
  (\bibinfo {year} {1993})}\BibitemShut {NoStop}%
\bibitem [{\citenamefont {Bonjour}\ \emph {et~al.}(1999)\citenamefont
  {Bonjour}, \citenamefont {Charmousis},\ and\ \citenamefont
  {Gregory}}]{Bonjour:1999kz}%
  \BibitemOpen
  \bibfield  {author} {\bibinfo {author} {\bibfnamefont {Filipe}\ \bibnamefont
  {Bonjour}}, \bibinfo {author} {\bibfnamefont {Christos}\ \bibnamefont
  {Charmousis}}, \ and\ \bibinfo {author} {\bibfnamefont {Ruth}\ \bibnamefont
  {Gregory}},\ }\bibfield  {title} {\enquote {\bibinfo {title} {{Thick domain
  wall universes}},}\ }\href {\doibase 10.1088/0264-9381/16/7/318} {\bibfield
  {journal} {\bibinfo  {journal} {Class. Quant. Grav.}\ }\textbf {\bibinfo
  {volume} {16}},\ \bibinfo {pages} {2427--2445} (\bibinfo {year} {1999})},\
  \Eprint {http://arxiv.org/abs/gr-qc/9902081} {arXiv:gr-qc/9902081 [gr-qc]}
  \BibitemShut {NoStop}%
\end{thebibliography}%

\end{document}